\documentclass[journal]{IEEEtran}

\usepackage{graphicx}
\usepackage{url}

\usepackage{amsmath} 
\usepackage{color,soul}
\usepackage{booktabs}

\ifCLASSINFOpdf

\begin{document}
%
\title{Programmable Crossbar Quantum-dot Cellular Automata Circuits}
%
%
%
	\author{Vicky~S.~Kalogeiton,~\IEEEmembership{Member,~IEEE}
		Dim P. Papadopoulos,~\IEEEmembership{Member,~IEEE}
		Orestis~Liolis,~\IEEEmembership{Member,~IEEE}
		Vassilios~A.~Mardiris,~\IEEEmembership{Member,~IEEE}
		Georgios~Ch.~Sirakoulis,~\IEEEmembership{Member,~IEEE}
		and~Ioannis~G.~Karafyllidis,~\IEEEmembership{Member,~IEEE}
		\thanks{Vicky~S.~Kalogeiton, Dim P. Papadopoulos, Orestis~Liolis, Georgios~Ch.~Sirakoulis and~Ioannis~G.~Karafyllidis are with the Department of Electrical and Computer Engineering,
			Democritus University of Thrace, Xanthi GR--67100, Greece.}
		\thanks{Vassilios~A.~Mardiris is with the Technological Educational Institute of Kavala, Kavala GR--65404, Greece.}        
	}


\maketitle

\begin{abstract}
Quantum-dot fabrication and characterization is a well-established technology, which is used in photonics, quantum optics and nanoelectronics.  Four quantum-dots placed at the corners of a square form a unit cell, which can hold a bit of information and serve as a basis for Quantum-dot Cellular Automata (QCA) nanoelectronic circuits. Although several basic QCA circuits have been designed, fabricated and tested, proving that quantum-dots can form functional, fast and low-power nanoelectornic circuits, QCA nanoelectronics still remain at its infancy. One of the reasons for this is the lack of design automation tools, which will facilitate the systematic design of large QCA circuits that contemporary applications demand.  Here we present novel, programmable QCA circuits, which are based on crossbar architecture. These circuits can be programmed to implement any Boolean function in analogy to CMOS FPGAs and open the road that will lead to full design automation of QCA nanoelectronic circuits. Using this architecture we designed and simulated QCA circuits that proved to be area efficient, stable and reliable. 
\end{abstract}


\IEEEpeerreviewmaketitle

\section{Introduction}
\IEEEPARstart{Q}{uantum}-dot Cellular Automata (QCA) is a promising nanoelectronic technology, in which information is stored as configurations of electron pairs in coupled quantum dot arrays. In QCA circuits these arrays are used \cite{Amlani00} to implement Boolean logic functions. More specifically, taking advantage of the quantum mechanical effects, QCA significantly reduce the size of digital circuits and operate at high speeds in very low power levels. QCA cells change their states due to interactions with neighboring cells via electrostatic or magnetic fields. Consequently, QCA, instead of using ranges of voltages or currents to represent binary values, they use the position of electrons in quantum dots. QCA integrated circuits have been implemented in densities up 1012 cells/cm$^2$ and the circuit switching frequency can be close to a terahertz \cite{ITRS,Orlov}.

Since 1993 when QCA were introduced by Lent et al. \cite{Lent93}, several logic gates, circuits and design rules \cite{Liu11,Kim:2006:QCA:1184198.1184358,Janez2012501} have been proposed such as the binary wire \cite{Lent93a}, the majority gate, AND, OR, NOT and XOR gates \cite{Tougaw94}, bit-serial adder, full adder \cite{Wang,Kim,Cho07}, multiplier \cite{Cho09}, multiplexer \cite{Vankamamidi,Mardiris09,Mardiris10}, flip-flop \cite{Huang,Momenzadeh,Shamsabadi}, arithmetic logic units (ALU) \cite{Niemier,Teja} and serial or parallel memories \cite{Vankamamidi08,Vankamamidi05,Taskin}. QCA circuits are generally stable, very fast and they consume very small amounts of energy, but the lack of design automation tools that will enable the design and simulation of large circuits and the lack of a scalable and modular architecture that will facilitate circuit fabrication do not allow the full development of this promising nanoelectronic technology. Furthermore, programmable pre-fabricated QCA circuits, such as microelectronic FPGA circuits, are expected to boost the use and applications of such circuits published in the literature.  

In this paper, we propose a novel design method of implementing Boolean functions using programmable QCA crossbar circuits. Crossings of horizontal and vertical nano-wire lines form a crossbar, which is considered as one of the most promising solutions for nanoelectronic circuit architectures \cite{Heath98}, because of its fabrication simplicity and the inherent redundancy, which supports defect tolerance \cite{Snider,Chen,DeHon,vourkas,Graunke05}. Its favorable properties include a periodic geometry, straightforward fabrication procedures, and a very compact definition of devices and interconnections, facilitating large scale fabrication and ultra high device density \cite{Heath98}. In the architecture proposed here, a QCA logic gate is formed at each cross-point of the crossbar. The logic gate can be programmed to operate as an OR, AND or NOT gate. These programmable gates form a universal Boolean set and any Boolean logic function can be implemented using this architecture, leading to QCA circuits that can execute any computation task. 

Furthermore, in order to provide designers with as much as possible flexibility a detailed methodology is introduced to enable the robust and efficient design of the corresponding QCA circuits. The proposed method takes into account the input/outputs as well as the considered programming lines of the crossbar architecture to implement Boolean functions and standard QCA circuits with the help of QCADesigner simulation tool \cite{citeulike:344490}. The timing issues of QCA cells and gates are successfully handled in every case with a cascadable, easy to follow way. The resulting QCA circuits use less unit cells and occupy less area compared to the state of the art QCA circuits.

The structure of the paper is as follows: In Section \ref{sec:qca1}, the necessary background for the QCA circuits is provided. In Section \ref{sec:set}, the proposed programmable basic logic QCA gates, i.e. AND, OR and NOT, that are formed at line crossings are presented. In Section \ref{sec:method}, the design method of programmable crossbar QCA circuits is described analytically. In Section \ref{sec:exper} the proposed method is applied and evaluated by implementing several Boolean functions. The corresponding results are discussed and compared with well-known QCA circuits published in the literature. Finally, conclusions and future perspectives are drawn in Section \ref{sec:conclusions}.

\section{QCA Preliminaries}
\label{sec:qca1}
\begin{figure}[hbp]
\centering
\includegraphics[width=0.45\textwidth]{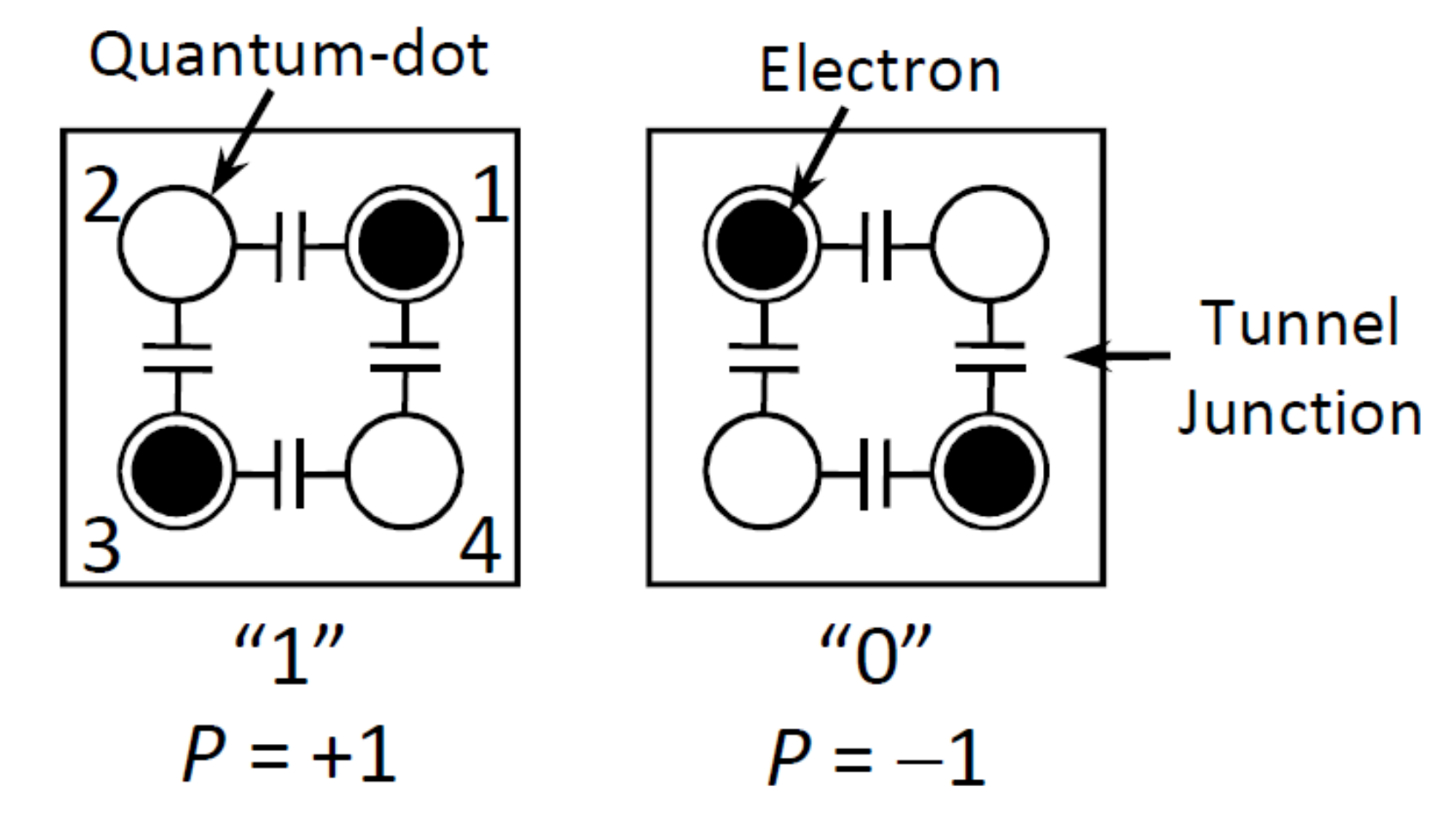}
\caption{The basic QCA cell.}
\label{QCAcell}
\end{figure}

The basic building block of QCA devices is the QCA cell presented in Fig. \ref{QCAcell}. It consists of four quantum dots in a square array coupled by tunnel barriers. The physical mechanisms for interactions between dots are the Coulomb interactions and quantum-mechanical tunneling. Electrons are able to tunnel between the dots, but cannot leave the cell. If two mobile electrons are placed in the cell, in the ground state and in absence of external electrostatic influence, Coulomb repulsion will force the electrons to dots on opposite corners \cite{Lent93}. The two possible charge configurations are presented in Fig. \ref{QCAcell} and are corresponding to binary ``1'' and ``0''.

For an isolated cell, the two polarization states are energetically equivalent. However, when a neighboring QCA cell is near, the equivalency breaks and only one of the two states becomes the cell ground state \cite{Lent93}. Polarization, $P$ measures the extent to which the charge distribution is aligned along one of the diagonal axes. If we label the four dots from 1 to 4 anti-clockwise starting from the upper right dot of the cell, and assign $\rho_i$ as the electron density of the $i^{th}$ dot, $P$ is defined as:

\begin{equation}
	P = \frac{(\rho_1+\rho_3)-(\rho_2+\rho_4)}{\rho_1+\rho_2+\rho_3+\rho_4} 
\label{eq4}
\end{equation}

The stability of QCA circuit is based on the assumption that the system falls to the ground state every time is excited by the inputs. However, this is not always guaranteed so in this case the system settles in a metastable state, affecting the functionality of the design. This problem can be solved by adiabatic switching \cite{Amlani97}. In adiabatic switching the system is always kept in its instantaneous ground state by using a clocking scheme sequence of four periodic phases. QCA cells receive the clock signal through an electric field which can raise or lower the tunneling barrier between dots inside the cell. When the barrier is low the electrons can move from one dot to another according to the overall external electrostatic influence. In case of high barriers the electrons are locked inside the dots so the external fields can not change the state of the cell. The adiabatic switching clocking scheme consists of four phases: $Switch$, $Hold$, $Release$ and $Relax$.

The adiabatic switching clocking scheme is implemented by applying four clock signals to the QCA circuit in order to control the clocking phase of each QCA cell in the circuit. In order to use the adiabatic switching clocking scheme the QCA circuit must be partitioned into clocking zones in such a manner so that all cells in a clocking zone are controlled by the same clock. The clocking zone partitioning is a crucial factor of the QCA design, because the order of appearance of the four clocking zones in the circuit controls the flow direction of the signals inside the QCA circuit.

At the beginning of the $Switch$ phase, the QCA cells in the zone are unpolarized since the cell tunneling barriers are low. During the switch phase the barriers are raised and the QCA cells become polarized according to the state of the cells that drive the zone. The driver cells must belong to a different clocking zone and specifically at the hold phase (90$^\circ$ phase difference, leading). $Switch$ phase is the clock phase that the actual computation (or switching) occurs. At the end of switch phase, barriers are high enough to block any electron tunneling and this locks the state of the cell. Next phase is the $Hold$ phase. During this phase the barriers remain high so the zone outputs can drive the inputs of the next clocking zone sub-circuit. The next zone (which is driven from our reference zone) must be at switch state (90$^\circ$ phase difference, leading). At the $Release$ phase, barriers are gradually lowered and finally cells are allowed to relax to an unpolarized state. Finally, during the fourth clock phase, the $Relax$ phase, cell barriers remain lowered and cells remain in an unpolarized state.

\begin{figure}[hbp]
\centering
\includegraphics[width=0.45\textwidth]{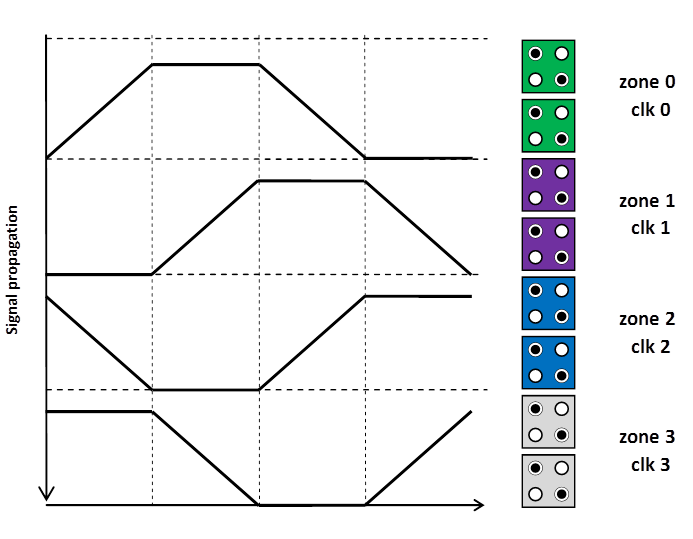}
\caption{Adiabatic clocking scheme application example on a binary wire.}
\label{QCAclock}
\end{figure}

Fig. \ref{QCAclock} presents an adiabatic clocking scheme application example on a binary wire. In the example the state of the two upper QCA cells constitute zone 0 is propagated gradually to the bottom zone 3 cells, according to the clocking mechanism presented above. The diagram in the figure shows the clocking phases of each zone for the duration of the signal propagation.

A row of QCA cells acts like a wire usually called binary wire \cite{Lent93a}. In QCA circuit designs of binary wires, inverters and three-input majority gates are the fundamental parts. The inverters are constructed with a fork structure and the majority gates with a cross structure \cite{Tougaw94}. Coplanar binary wire crossovers can also be implemented in QCA designs which is a very useful technique for designing more realizable circuits \cite{Tougaw94}.

\section{Programmable logic Gates for QCA crossbar circuits}
\label{sec:set}

\begin{figure}[htbp]
\centering
\includegraphics[width=0.3\textwidth]{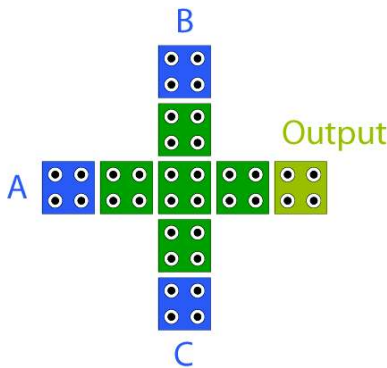}
\caption{The majority logic QCA Gate. Using one of its inputs as a program line, the majority gate becomes a programmable AND/OR gate.}
\label{QCAmaj}
\end{figure}

Crossings of ``horizontal'' and ``vertical'' QCA wires form programmable majority gates which have been fabricated and tested \cite{Orlov}. Fig. \ref{QCAmaj} shows a majority logic QCA gate formed at such a crossing. The gate comprises five cells in each direction. The three inputs to this gate are the top, left and bottom cells, A, B and C and the output is the right cell. The output is equal to the majority of the input states. The central cell, which is the one that performs the calculations is called the ``device cell'' and in every case is in the same state as the output. One of the three inputs of the majority gate can be used as a ``program line''. Let us assume, that the program line comes in from the top input cell, namely B, (of course any one of the input cells can serve as a program line due to symmetry), and the rest of the two inputs are free to be set at any state. In such a case, one can see that if the program line state is set to one, the majority gate actually performs the OR Boolean logic, while when the program line is set to zero, the majority gate becomes an AND gate. Therefore, the majority gate can be programmed to function as an OR or an AND gate by setting the state of any one of its inputs to logic one or zero and keeping it constant during the operation of the gate. Although QCA wire crossings can be programmed to function as AND or OR gates, they cannot be programmed to operate as NOT gates by setting any two inputs to any logic value. 

\begin{figure}[htbp]
\centering
\includegraphics[width=0.3\textwidth]{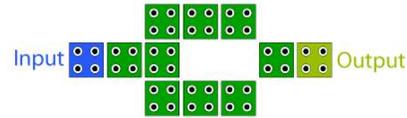}
\caption{The standard QCA NOT gate design.}
\label{QCAnot}
\end{figure}

The frequently used QCA NOT gate (Fig. \ref{QCAnot}) is based on the fact that the input signal comes from the left of a QCA wire and splits into two parallel offset wires. The signal is inverted at the right end of the circuit forming thus a NOT gate. As it can be easily observed, such a gate is not applicable to the crossbar architecture, due to the fact that its structure is different from the one the crossbar demands.

\begin{figure}[htbp]
\centering
\includegraphics[width=0.3\textwidth]{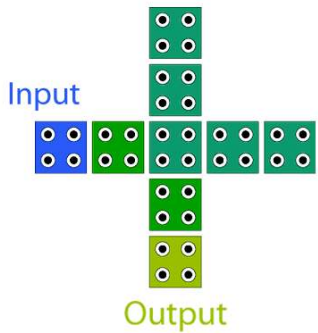}
\caption{The proposed QCA NOT gate in crossbar architecture.}
\label{QCAnotb}
\end{figure}

Consequently, in the context of this work, a different implementation of a QCA NOT gate is proposed. As demonstrated in Fig. \ref{QCAnotb}, the proposed NOT gate is suitable for the design of QCA circuits in crossbar architecture, given that it can be implemented in a cross point of the crossbar. In particular, the proposed NOT gate has the same cells' topology as the pre-mentioned majority logic QCA gate, namely it comprises five cells in each direction. Among all nine cells of the gate, only four of them are useful for the signal inversion, since these are the ones that transfer the signal. These cells shown in Fig. \ref{QCAnotb} with green color, are triggered by the same clock, while the rest five cells (blue color) are triggered by another clock so that these two clocks do not overlap. Due to symmetry, the input signal may come in from any side of the gate; however, let us assume that the left side of the gate serves as input wire. In this case, the inverted signal exits the gate from the top or bottom cell. In other words, the incoming and the outgoing wires should be perpendicular to each other. This means that the proposed NOT gate is based on the diagonal orientation of two cells, namely in Fig. \ref{QCAnotb} the last from the left cell and the upper cell of the incoming and outgoing wire, respectively. These cells in diagonal orientation, due to local Coulomb interactions, tend to be aligned in opposite polarization directions, in the same way that two rotated cells in a horizontal orientation do. Thus, the inverted signal exits the gate from the output cell $Output$ and following the principle of minimum energy corresponds to the stable condition of the gate. As a result, the proposed QCA inverter is able to be applied in a crossbar as long as its cells follow the aforementioned structure and timing issues. In specific, we take advantage of the diagonal orientation of the two cells ``last from the left'' of the incoming wire and ``upper cell'' of the outgoing wire to obtain inversion. This will invert a ``0'' to ``1'' with high reliability and, analogously, taking advantage of the rest of the inserted column cells this will also invert a ``1'' to ``0'' with the same high reliability and stability. In terms of polarization of of the NOT gate cells in the QCA crossbar, the resulting charge distribution based on QCADesigner simulation results from the energy distribution of the QCA cells and the different phases of the two clocks established in the proposed NOT gate. In specific, the stability of the proposed gate is provided by the Coherence Vector model \cite{citeulike:344490} measurements which calculate the state of the cells based on the accumulated kink energies. In all possible combinations of polarizations that take into account the proposed different clock zones and the number of cells between cross points, the Coherence Vector machine produces kink energies for the inverted cells that keep the polarization steady in every case.

\begin{figure}[htbp]
\centering
\includegraphics[width=0.3\textwidth]{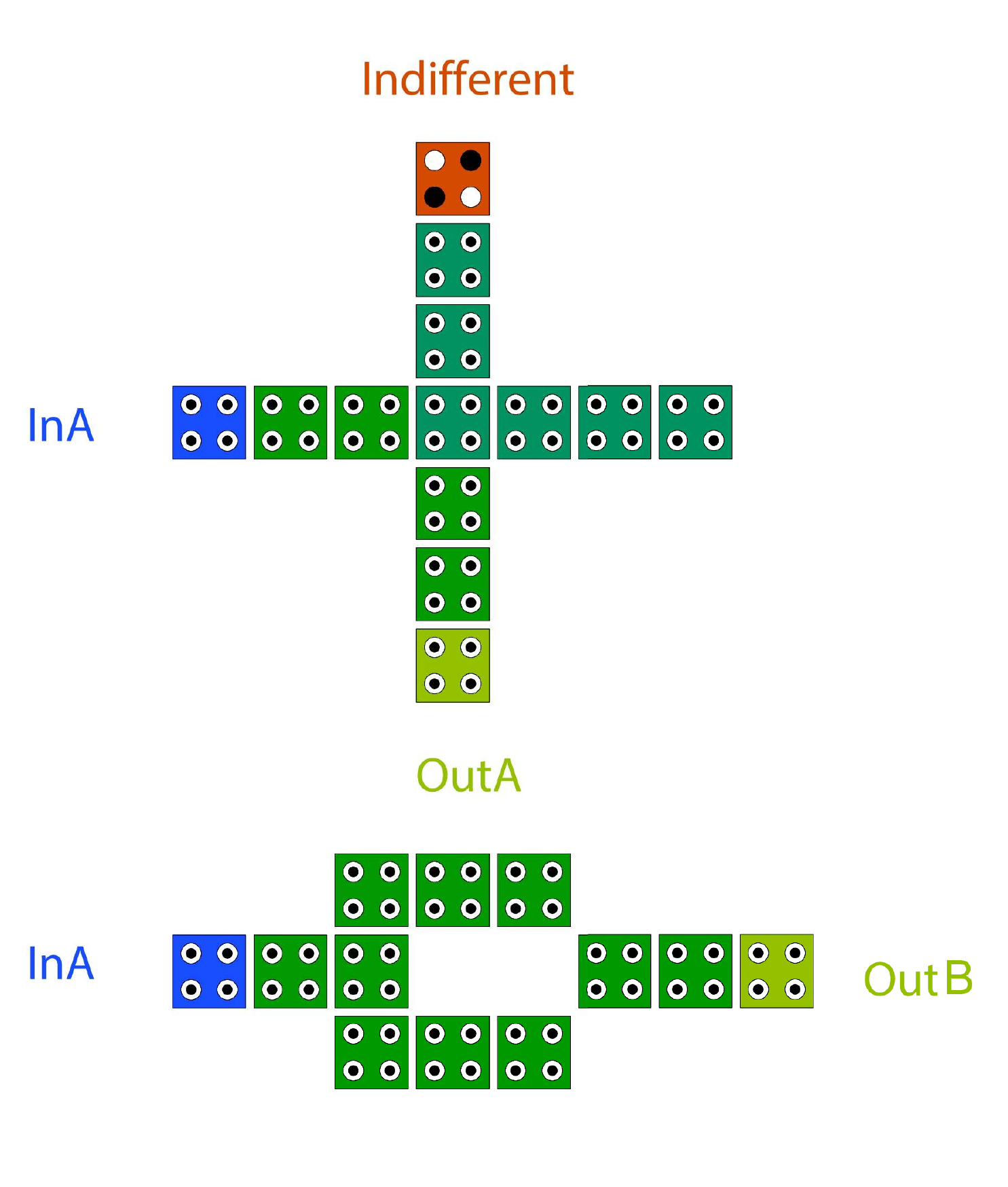}
\caption{Standard QCA NOT gate vs our proposed QCA NOT gate.}
\label{NOTgates}
\end{figure}

\begin{figure}[htbp]
\centering
\includegraphics[width=0.3\textwidth]{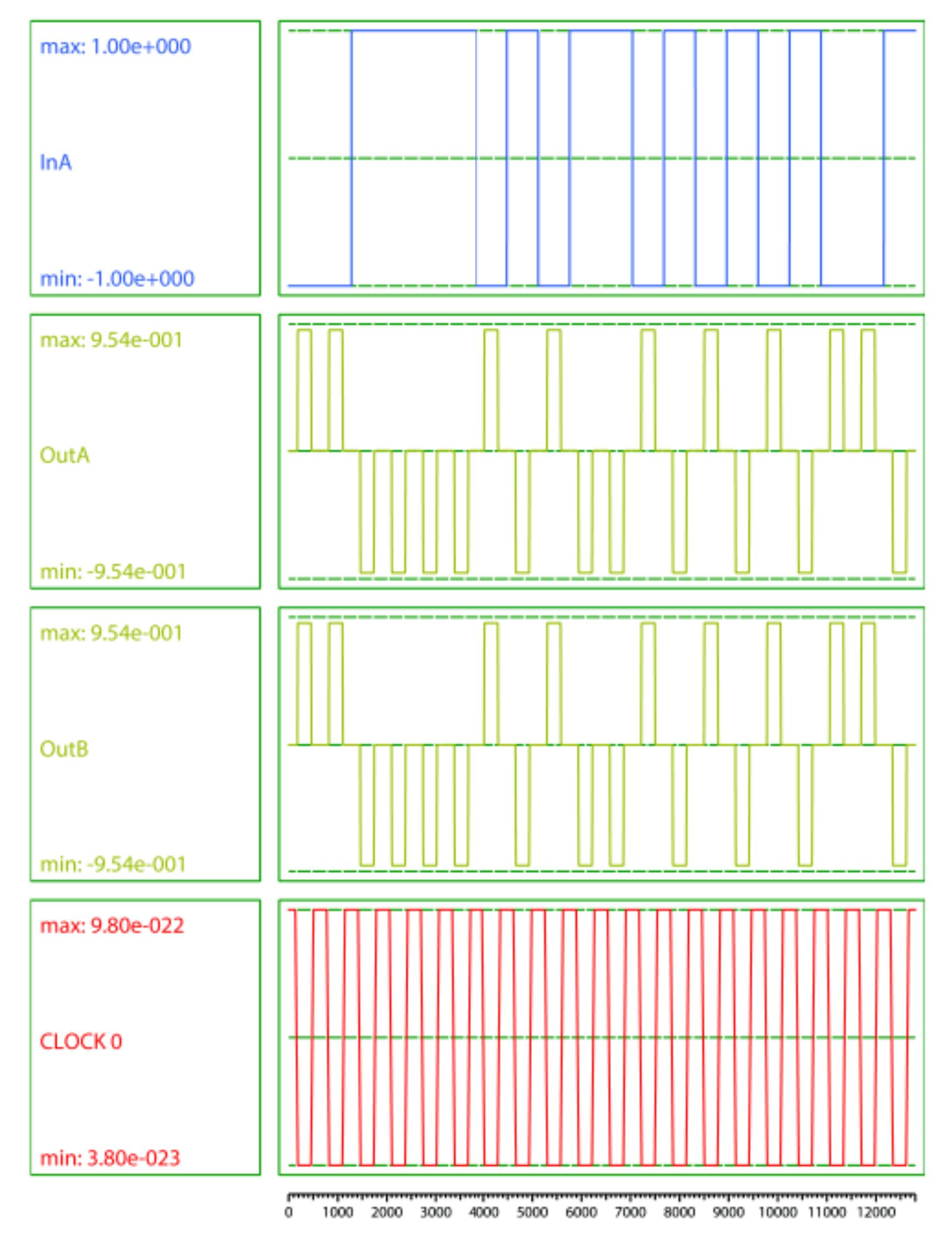}
\caption{Same inputs are inserted to the two different QCA NOT gates. The simulation results indicate that these two different implementations appear to be equally stable and reliable.}
\label{NOTgatesresults}
\end{figure}

More specifically, to verify the functionality of the proposed in this paper NOT gate we compared the results of the proposed NOT gate to the ones of the standard QCA NOT gate in terms of stability and robustness. Therefore, Fig.~\ref{NOTgates} illustrates two simple circuits, each of which contains only one of the pre-mentioned QCA NOT gates and the same input is inserted into both of them. The results of both implementations of the QCA NOT gates are demonstrated in Fig.~\ref{NOTgatesresults}. By observing these results, it is obvious that the proposed QCA NOT gate is able to invert both states with high reliability and stability. 

\begin{figure}[htbp]
\centering
\includegraphics[trim={2.5cm 5cm 2cm 4.25cm}, clip=true, width=0.2\textwidth]{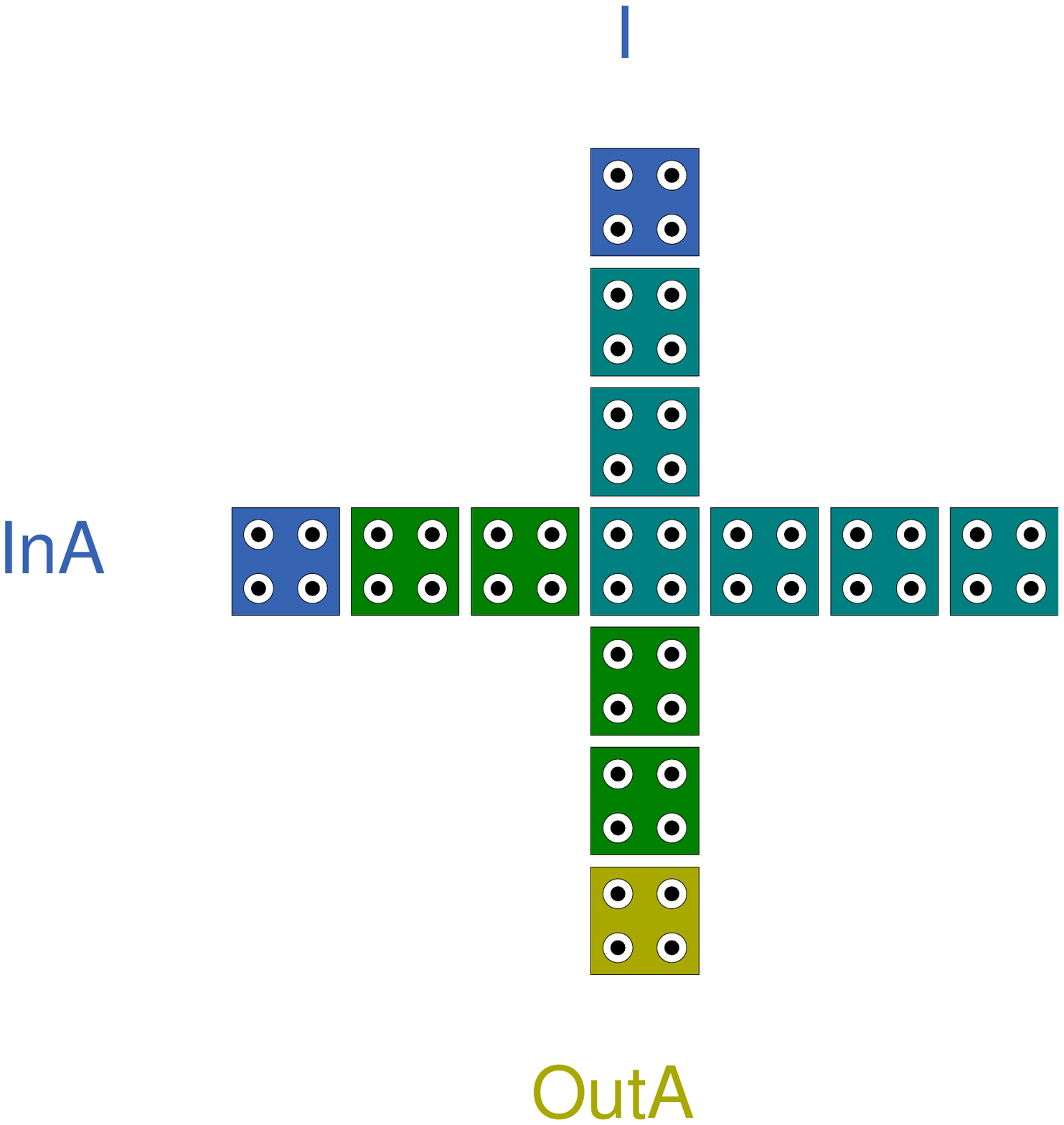}
\caption{Input and output signals of the proposed QCA NOT gate implemented in the cross point of the crossbar.}
\label{indepent}
\end{figure}

\begin{figure}[htbp]
\centering
\includegraphics[trim={2.25cm 2.5cm 1.75cm 3cm}, clip=true, width=0.3\textwidth]{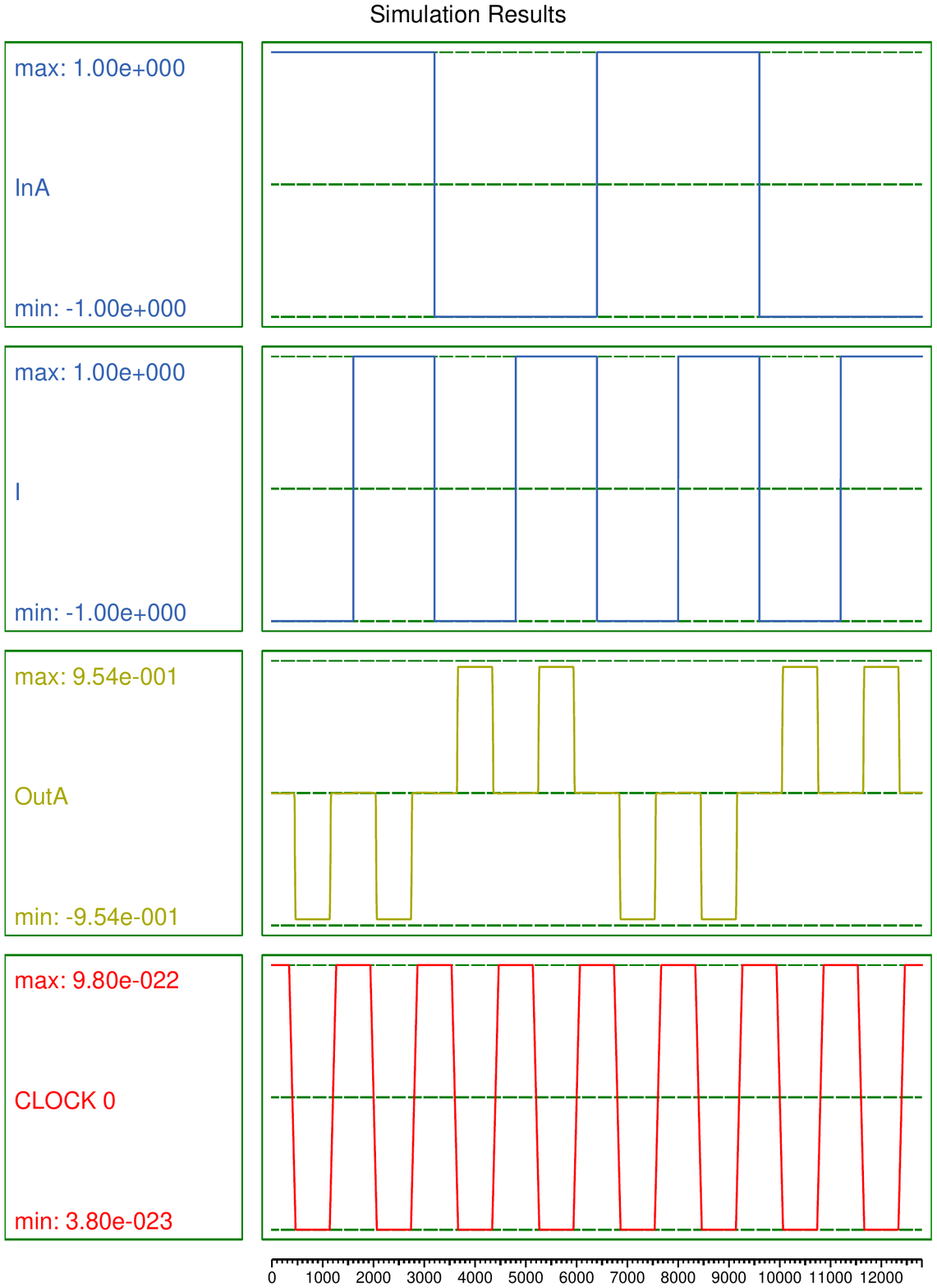}
\caption{Simulation results for all possible values of the input signal \texttt{InA}, signal \texttt{I} and the corresponding output \texttt{OutA} of the proposed QCA NOT gate. In all cases NOT gate succeeds to invert the input signal \texttt{InA} with high reliability and stability.}
\label{newNOTresults}
\end{figure}

Moreover, in Fig.~\ref{indepent} the functionality of the proposed NOT gate is analysed for all possible input combinations combined with all possible values of the indifferent signal \texttt{I}. The results are shown shown in Fig. \ref{newNOTresults}. For all values of signal \texttt{I}, the input signal \texttt{InA} is always inverted in the output signal \texttt{OutA}, which is located at the lower cell of the crossbar design. The successful inversion of input \texttt{InA} in all possible cases, no matter which values take input signal \texttt{I} as presented in Fig. \ref{newNOTresults}, corroborate the high reliability and stability of the inversion succeeded by the proposed NOT gate. We would like to emphasize that the only condition for this inversion is the proper timing of the input--output related cells shown in Fig. \ref{QCAnotb} with green color, which are triggered by the same clock, while the rest five cells (blue color) of signal \texttt{I} and the indifferent most right cells are triggered by another clock. The  two clocks do not overlap and this timing asymmetry compensates for the spatial asymmetry of our NOT gate, resulting in the inversion of both ``1'' and ``0'' with the same reliability.

\section{Design Methodology and Programming of Crossbar QCA Circuits}
\label{sec:method}
In this section, we will describe the method for designing and programming crossbar QCA circuits that implement any given Boolean functions. The programming of crossing points to form the basic logic gates described in the previous section must follow certain rules that will be described later on. The crossbar architecture demands on one hand an appropriate location for each gate of the circuit, corresponding to suitable connection between the gates of the circuit and on the other hand a proper timing, meaning that each cell should be triggered by a specific clock wave at the correct time phase. We want to develop a design methodology for programmable QCA crossbar circuits in analogy to the CMOS FPGA circuits. However, the combination of the QCA circuits with the crossbar architecture is not only a novel task but also a quite promising one, because of the resulting stability in conjunction with adaptability and regularity.

\begin{figure}[hbp]
\centering
\includegraphics[width=0.45\textwidth]{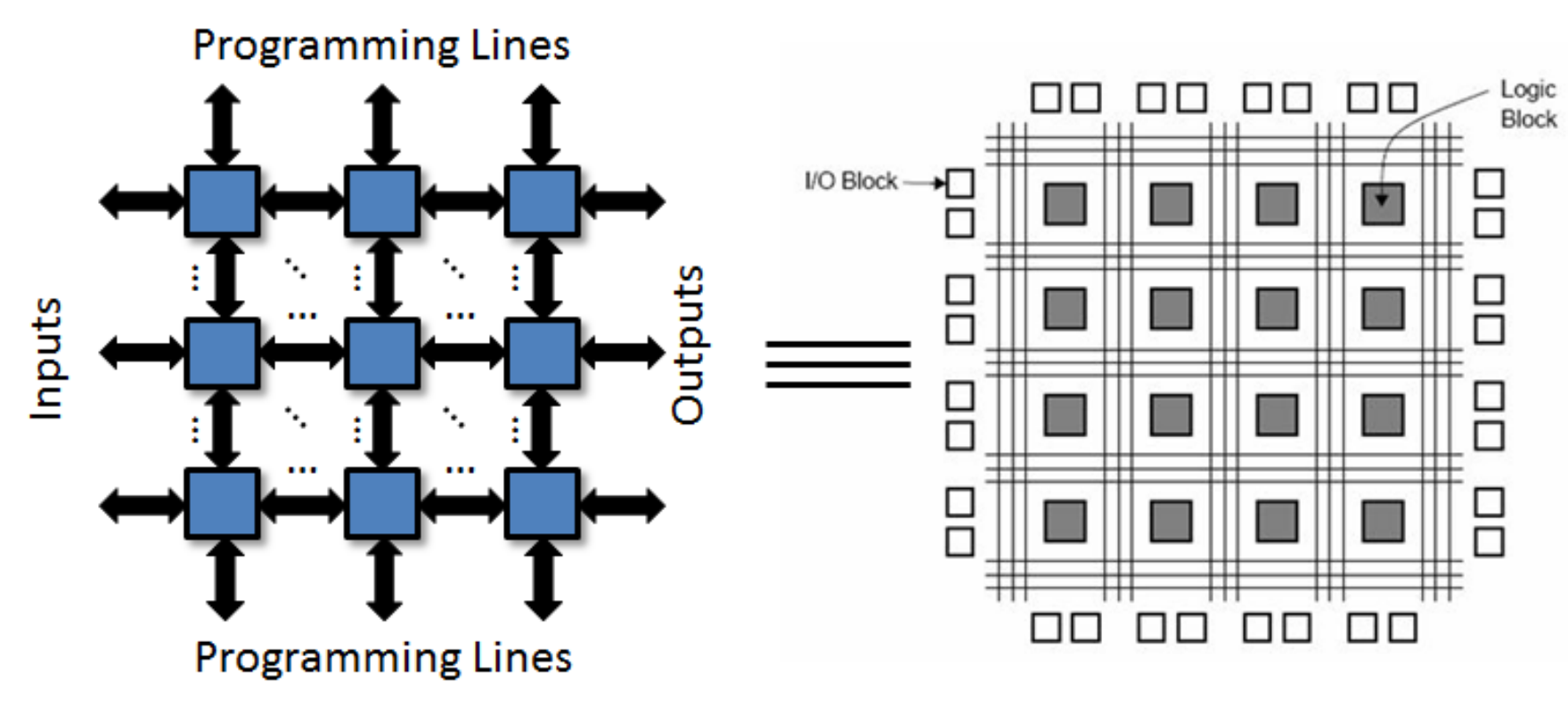}
\caption{The proposed programmable QCA crossbar architecture and its analogy to the FPGA CMOS architecture.}
\label{QCAcross}
\end{figure}

The following design rules should be followed:

\underline{Inputs/ Outputs:} As shown in the crossbar architecture of Fig. \ref{QCAcross}, each input signal should be applied to the one side of the crossbar, while the desired output should be extracted from the opposite side. This way there will be no conflict between the inputs and outputs of the circuit, since the incoming and outgoing signals will be in different sides. In the cases treated in this work, without loss of generality, the input cells are the ones on the left side of the crossbar while the output ones are located on the right side. For the sake of brevity, the left side of the crossbar will be called \textit{input side}, while the right one \textit{output side}.

. 
\underline{Program Cells:} The cells located at the top and bottom of the crossbar, are the cells used to program the circuit. These cells are ``anchored'', i.e. their value is constant and is determined (by the designer) during the design process of the circuit. This provides the designer the ability to program the circuit so that it can execute any Boolean function. If there is no need of a specific value, the value of these program cells is indifferent and does not affect the function of the rest of the circuit. A characteristic example of this indifferent value is demonstrated in Fig.~\ref{QCAnotb}, where the proposed QCA NOT gate functions correctly and properly despite of the value of the cell called \textit{Indifferent}. Otherwise, this desired value enters the circuit through these cells. At this point it should be noticed that the value inserted from the top/ bottom of the crossbar remains constant during the circuit operation and, therefore, it is not handled as a separate input. Thus, the top and bottom cells will referred to as \textit{program lines}. 

\underline{Branches:} The circuit should not have any branches, given that the crossbar architecture does not support them.

\underline{Basic Gates:} Each circuit should contain only the basic gates already analyzed in Section \ref{sec:set}. That means that every other gate i.e. XOR gate of the circuit should be constructed using the basic gates that are considered to be the basic modules of the circuit. 

\underline{Gate Placement:} The cross point where each gate is located into the crossbar is of great importance. Each OR and AND gate operation stems from the majority logic gate. As a short reminder the last one presupposes three input signals. In particular, the programming input signal is different in both gates and is essential in order for them to operate appropriately. However, this signal can not be considered as an input. Hence, it should be inserted to the main circuit through the program line of the crossbar, as described earlier in this section. This fact implies that the implementation of these gates is only available next to the \textit{program lines}, namely next to the bottom/ top side. The NOT gate can be placed wherever into the circuit, since it does not necessitate any extra input signal.

After dealing with the aforementioned parameters that the designer should take into account, the next step of the proposed design method refers to the dimensions of the resulting QCA circuit in conjunction with crossbar architecture. These dimensions can be defined according to the following design rules: 

\begin{enumerate}
\item The number of crossbar horizontal lines $N_{lines}$ should be at least equal to the number of the circuit inputs, i.e. $N_{inputs}$ and less or equal to the sum of the number of the circuit inputs and the number of the circuit NOT gates $N_{NOTgates}$ as follows: 

\begin{equation}
N_{inputs} \leq N_{lines} \leq N_{inputs} + N_{NOTgate}
\label{eq1}
\end{equation}

The first part of the inequality expresses the fact that each input needs at least one line in order to be inserted to the circuit. With regard to the second part of the inequality, this stems from the fact that each NOT gate of the circuit demands two lines to be implemented in the crossbar. More specifically, the first line is occupied by the initial signal while the second one is occupied by the inverted one. Equality arrives when the inverted signal can be produced in an existing line, given that the signal that already occupied that specific line is no longer useful for the rest of the circuit; therefore it can be replaced by the inverted one.

\item The number of crossbar vertical columns $N_{columns}$, corresponding to the program lines, is directly associated with the number of the circuit stages. The term ``stage'' refers to the minimum number of columns $N_{columns}$ needed in order for all the gates of the circuit to be designed. Introspectively, the determination of the $N_{stages}$ emanates from the route that each input signal follows as follows: for all input signals of the circuit the number of majority logic gates, i.e. AND, OR and majority gates, that each signal passes though is calculated. In this way, the $N_{stages}$ corresponds to the largest number of them, given that it indicates the minimum $N_{columns}$ needed so that all signals exit the circuit. As a result, when a majority gate's input signal is another gate's output one, then these gates are located in separate columns, while others that are able to function independently can be incorporated into a single column, constituting a stage. At this point, it is worth mentioning that the NOT gates, through which a signal passes, are not taken into account, since a NOT gate, when available, can be implemented into a single column and therefore, its input signal is able to enter two different gates at a single column.

However, as already mentioned, there are cases where this incorporation of gates in one column is not possible, e.g. the NOT gate demands two crossbar lines or even the incorporation of 3 AND gates in one column is not feasible, given that they demand program cells that according to the aforementioned rule are located to the top/bottom of the crossbar. Consequently, in worst case scenarios where no incorporation is achievable, each gate occupies one column and therefore, the maximum  number of columns needed equals to the number of the circuit gates $N_{gates}$. In that way, the crossbar $N_{columns}$ is described by the following inequality: 

\begin{equation}
N_{stages} \leq N_{columns} \leq N_{gates}
\label{eq2}
\end{equation}

\end{enumerate}

\begin{figure}[htbp]
\centering
\includegraphics[width=0.45\textwidth]{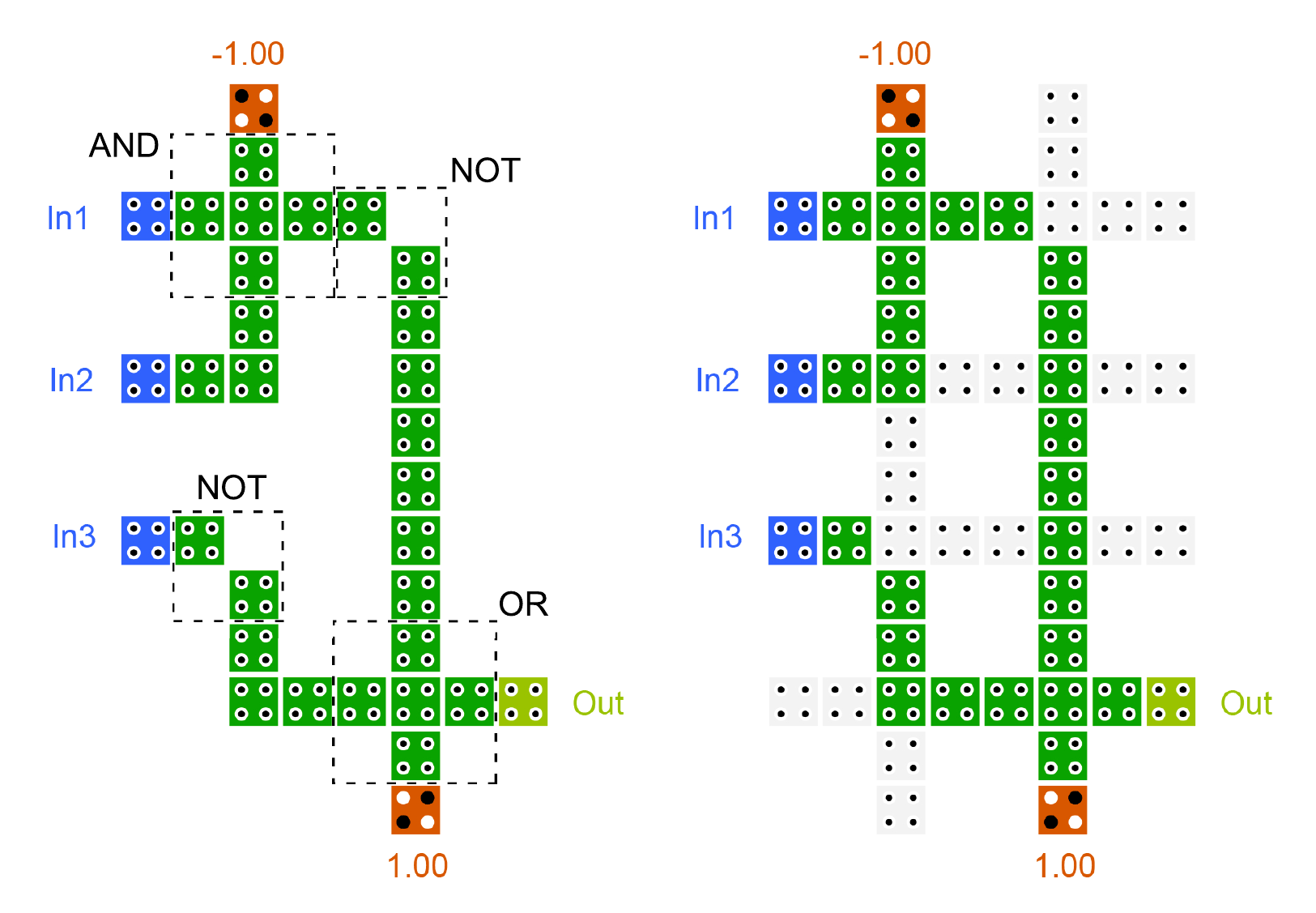}
\caption{Example of the proposed design rules when applied to a simple QCA circuit in a crossbar architecture.}
\label{Fig1}
\end{figure}

Fig.~\ref{Fig1} depicts the application of these two basic design rules for a Boolean circuit. For example, let us assume, a simple circuit with three inputs and four gates, two of which are NOT gates. $N_{lines}$ i.e. $4$ of the example circuit is greater than $N_{inputs}$ i.e. $3$, but less than the sum of $N_{inputs}$ and $N_{NOTgates}$, i.e. $4$. This arrives from the fact that each NOT gate occupies two lines, (please refer to Section \ref{sec:set}). However, in case of the first (top) NOT gate, there is an already available line while in the second one there is not. On the other hand, $N_{columns}$ is less than $N_{gates}$, i.e. $4$, and in particular is equal to the $N_{stages}$, i.e. $2$, since both the input signals $In1$ and $In2$ pass though one AND and one OR gate. Hence, the first stage contains the AND gate and one NOT gate, while the second NOT gate is incorporated in the same column with the OR gate, constituting the second stage.

The next step of the design procedure refers to the timing of each cell. The timing is determined and linked to the clock function. The clock signal for each single cell is quite crucial, given that the timing of the signal should be accurate in order for the circuit to operate properly. More specifically, the clock function should be configured in such a way so that: \textit{a)} each signal comes in to the corresponding cell at the right time and \textit{b)} the either no longer needed or indifferent signals do not affect the circuit function.

Each signal in a cell is triggered by a specific clock. Initially, input cells of the \textit{input side} are activated first, triggered by the first clock, namely \texttt{Clock0}. Then, all other cells connected to these input cells are triggered by the same clock (\texttt{Clock0}), creating in that way \textit{wires} of cells functioning under the same clock. Thus, the input signal is able to propagate from the input cell through the main circuit.

After that, the timing of the gates follows. The timing of each gate is essential and the cells of each OR and AND gate, namely of each majority logic gate, are preferably operating under the same clock. This clock should be the very next to the one operating the input cells of each gate. That does not imply that a gate is not able to function properly when both its input cells and the gate itself are being triggered at the same time. However, in this way, each signal enters the gate at the correct time and any error, resulting from a late arrival of a signal to the gate inputs, is eliminated/ evaded. The NOT gate cells are operating at the same time, since the output/inverted signal of the NOT gate should be clocked with its input signal. Consequently, when a signal comes in an AND or OR gate automatically changes the clock that its cell is being triggered by. In other words, the cells of these gates change their operating clock, making it the next in line, namely \texttt{Clock1}.

The design and timing of a gate is followed by the next step, in which the operating clocks of cells are defined. As expected when a signal passes through a few cells, it can be transferred without any alteration. However, this does not happen when the signal comes in through more than $6-7$ cells in a row. The heuristic numbers of 6 to 7 cells in each row and column before each cross-point has been proposed after numerous simulation tests that finally lead us to reliable and stable results in every examined QCA circuit design. Because of that, every $6-7$ cells is preferable for the signal to be propagated using the next clock, i.e. \texttt{Clock2}.  At this point it should be emphasized that the proposed upper heuristic limit of $6-7$ cells can be conditionally reduced. For example, in case of entering through a gate, it is possible for a signal to alter its clock earlier. In order for the gate to operate properly, all its input cells should be triggered by the same clock. That means that in cases where the signals arrive at a gate unsynchronized, the gate can not generate the expecting output. Thus, the wires of cells leading to the gate inputs should change their operating clock earlier, namely before they come into the gate, resulting in possible alternation of clocks before the given limit of $6-7$ cells.

Apart from the aforementioned wire length, another important factor regarding the operating clocks definition is that of the interaction of adjacent cells. In particular, an indifferent or no longer needed signal of a cell should not influence its adjacent cells. This is feasible by the proper timing of that cell, i.e. when the operating clock of that particular cell is set to be a following clock, or to be a clock that does not have any relevance with the clocking of this cell.

If, for example, a desired signal is propagated through the \texttt{Clock1} and meets in a cross-point an indifferent signal, then the clock signal should be set to be \texttt{Clock3}, given that these two clocks do not overlap.

\begin{figure}[htbp]
\centering
\includegraphics[width=0.45\textwidth]{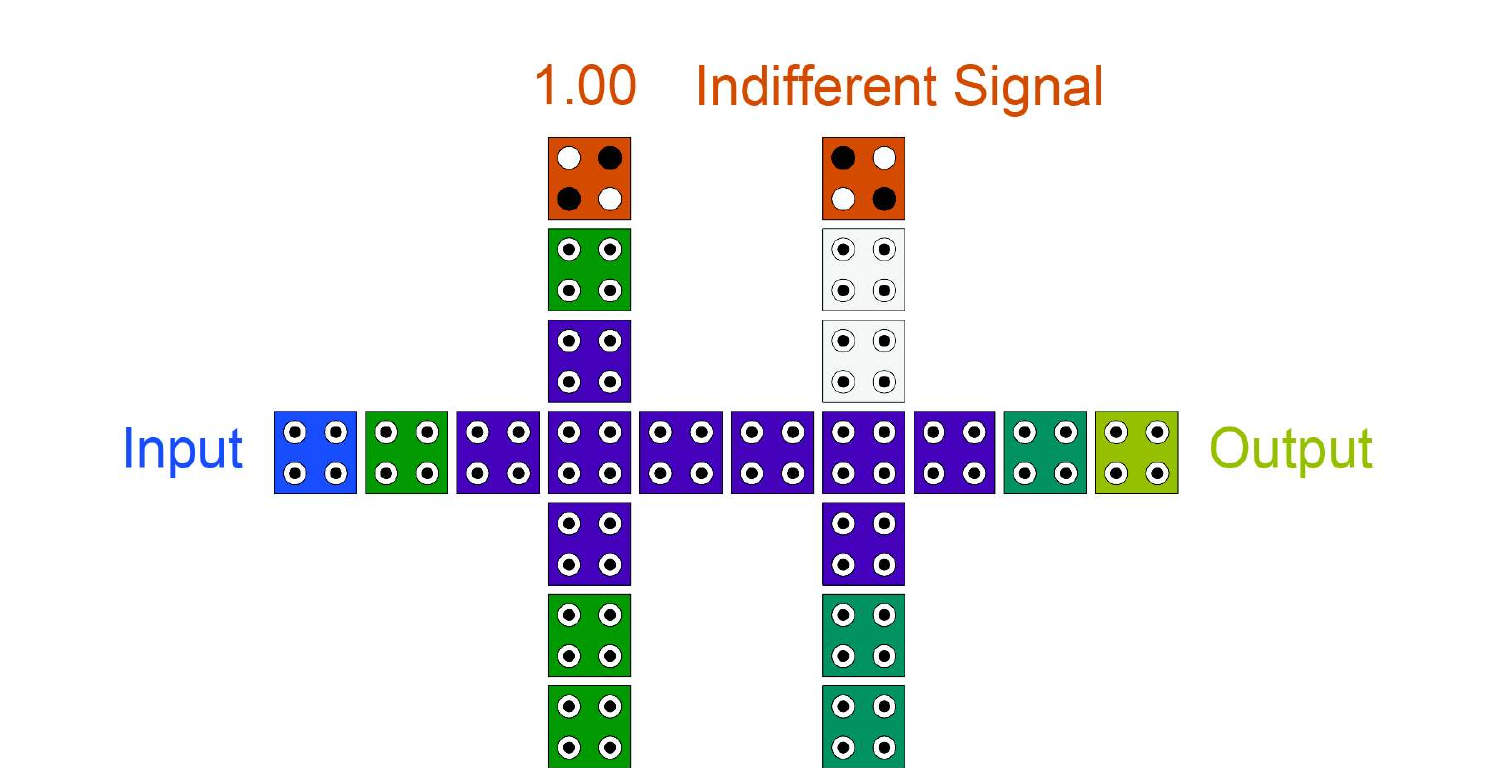}
\caption{Timing of cells and gates. \texttt{Clock0} is represented by green color, \texttt{Clock1} by purple, \texttt{Clock2} by blue and \texttt{Clock3} by white color, respectively. An AND gate is depicted and its result drives another part of the circuit, while it is also an output of the circuit.}
\label{Fig2}
\end{figure}

Fig.~\ref{Fig2} illustrates a characteristic example of the timing process described earlier. The input cells of the AND gate are triggered by the first clock, \texttt{Clock0} -all of them at the same time-, while the gate is operating at the next clock, namely \texttt{Clock1}. The wire that carries the desired signal is operating under \texttt{Clock1} for $7$ cells and then its clock is set to \texttt{Clock2}. The first from the right in Fig.~\ref{Fig2} program line carries an indifferent for the rest of the circuit signal and thus, its cells are functioning under \texttt{Clock3}, which is the second following in line clock after \texttt{Clock1}.

It should be noticed that the designer should repeat the above steps that refer to the timing of the cells and the gates as many times as needed, in order to successfully accomplish the desired design. 

\begin{figure}[htbp]
\centering
\includegraphics[width=0.45\textwidth]{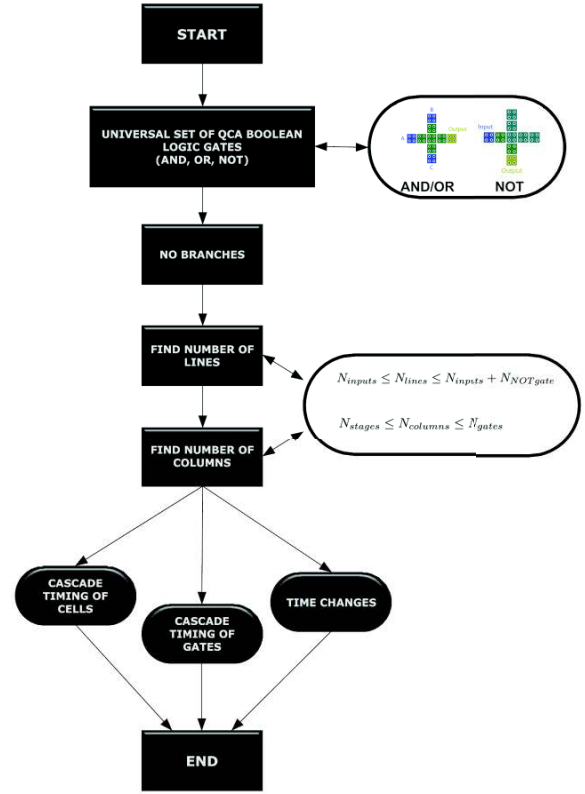}
\caption{The proposed design methodology in the form of block diagram}
\label{Flowchart}
\end{figure}

In conclusion, the proposed design method of crossbar QCA circuits introduced in this paper, as depicted in diagram form in Fig. \ref{Flowchart}, can be summed up as follows:

\begin{enumerate}
\item Provide a universal set of programmable QCA Boolean logic gates. 
\item Implement the corresponding circuit without branches with the help of the previous set.
\item Check $N_{inputs} \leq N_{lines} \leq N_{inputs} + N_{NOTgate}$.
\item Check $N_{stages} \leq N_{columns} \leq N_{gates}$.
\item Provide proper cascade timing of cells.
\item Provide proper cascade timing of gates.
\item Introduce timing changes where necessary.
\end{enumerate}

The cases of circuits with branches should be also examined. According to the universality of the provided Boolean set, it is expected for a circuit with branches to be redesigned as one without branches. This is accomplished as analytically described in \cite{haha}, where a detailed transformation of many cases using the majority logic gate is presented. Moreover, the proposed design method is based on a crossbar architecture where all the input signals enter the circuit/crossbar from the \textit{input side}. However, given this structure and the crossbar architecture, the design of a circuit where more than two AND or OR gates have to be calculated at the same stage simultaneously is not feasible. In more detail, in the \textit{input side} it is possible to have as many NOT gates as wanted but only two AND/ OR gates, since these gates demand a constant input from the \textit{program line}. This can be achieved without the limitations introduced earlier, namely if not only the inputs but also the constant values needed for the function of the AND/ OR gates are inserted from the \textit{input side} of the crossbar.

\section{Design Experiments}
\label{sec:exper}
\begin{figure}[htbp]
\centering
\includegraphics[width=0.45\textwidth]{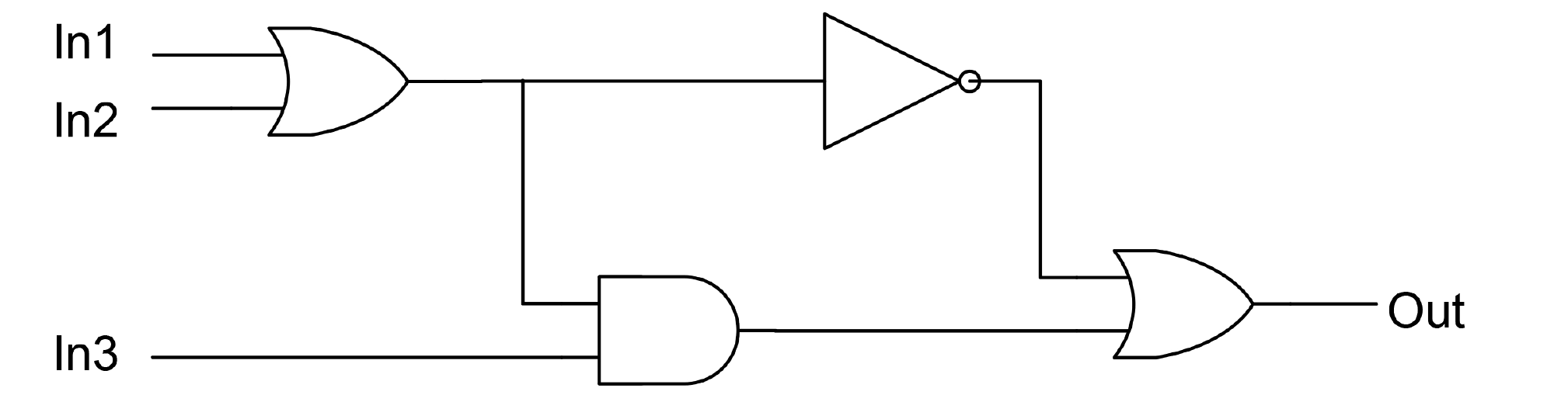}
\caption{Schematic design of first Boolean function \texttt{Out} = (\texttt{In1} + \texttt{In2})' + ((\texttt{In1} + \texttt{In2}) $\times${\texttt{In3}}) with three (3) inputs, one (1) output and four (4) logic gates}
\label{Fig3}
\end{figure}

\begin{figure}[htbp]
\centering
\includegraphics[width=0.45\textwidth]{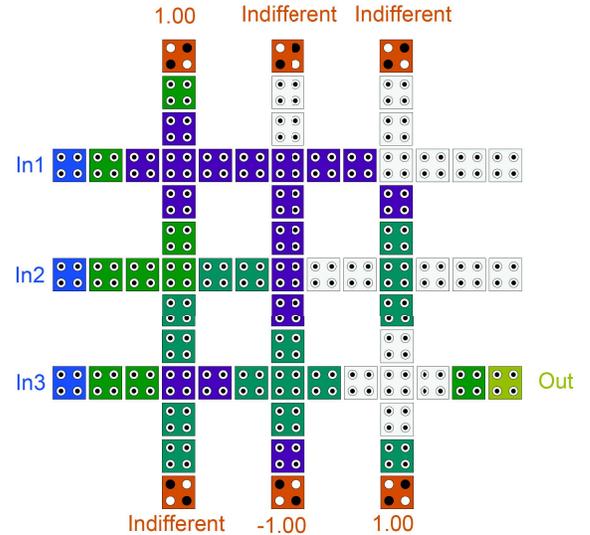}
\caption{The resulting crossbar QCA circuit corresponding to the Boolean function of Fig.~\ref{Fig3}.}
\label{Fig4}
\end{figure}

To evaluate the efficiency of the proposed method several design experiments were conducted. In particular, eight different Boolean functions of various complexity, which were successfully implemented in crossbar QCA circuits following the above method, are presented in this Section. The first two functions were selected to be simple so that the proposed method can be applied and described analytically. The next two circuits are the 2-to-1 multiplexer and the 4-to-1 multiplexer. Then, in order to reveal the ability of the proposed method to be applied to more complex circuits, two ISCAS bench mark circuits were selected (benchmark c17 from ISCAS'85 collection and benchmark s27 from ISCAS'89 collection) to be designed and simulated according to the proposed approach. Finally, the remaining two designs are the half-adder and the full-adder, that were selected to indicate the efficiency of the method introduced in this paper and to compare the circuits designed using this method with other QCA implementations found in literature. All circuits were designed and simulated using the QCADesigner, a design and simulation tool \cite{citeulike:344490}.
The first Boolean function consists of $3$ input and $1$ output signals as shown in Fig. \ref{Fig3}. Thus, according to eq. \ref{eq1} $N_{inputs} \leq N_{lines} \leq N_{inputs} + N_{NOTgate}$ and so $3 \leq N_{lines} \leq 4$. As far as the number of columns is concerned, these are determined from eq. \ref{eq2} as follows: $N_{stages} \leq N_{columns} \leq N_{gates}$ and so $3 \leq N_{columns} \leq 4$. Fig. \ref{Fig4} illustrates the corresponding QCA circuit. It can be easily observed that $ N_{lines} = 3$ and $N_{columns} = 3$. The numbers of the lines and the columns are the minimum ones according to eqs.~\ref{eq1} and ~\ref{eq2} respectively, since the inverted signal of the NOT gate can be transferred via the second, non-anymore-used line of the crossbar architecture and the gate itself is able to be incorporated in the same stage /column with the OR gate.

Regarding, the timing of the circuit, the input cells of the \textit{input side}, are triggered by the first clock, i.e. \texttt{Clock0}, that is distinguished by its green color. The first OR gate from the left, functions under \texttt{Clock1}, that is actually the next clock of the one its input cells are triggered. The output signal of the first OR gate is entered both to a NOT and an AND gate. This output signal is triggered by \texttt{Clock1}. Thus, as far as the AND gate is concerned, all its input signals should function under the same  clock, namely \texttt{Clock1}, and therefore the gate itself is triggered by the next clock, i.e. \texttt{Clock2}. \texttt{Clock1} and \texttt{Clock2} are presented with purple and blue-turquoise color, respectively. Finally, the second OR gate functions under \texttt{Clock3}, that is represented by white color. This clock is also the next in line from the one its input signals function in. This automatically entails that the output signal of NOT gate changes its clock to \texttt{Clock2}. The timing of the cells that are considered as indifferent follows the rules described earlier, meaning that the operation clock of these cells is set to be a following one or a clock that does not interfere with the clock of their adjacent useful cells. The input ($In1$, $In2$ and $In3$) and the output ($Out$) signals of the QCA circuit illustrated in Fig. \ref{Fig4} as well as the clock under which $Out$ operates are demonstrated in Fig. \ref{Graphs}(a). It should be noted that although the three inputs and the output are triggered by the same Clock, i.e. \texttt{Clock0}, the output delays four clock phases from the inputs. The obtained results are in accordance with the truth table of the logic circuit, proving the effectiveness of the proposed methodology.

\begin{figure}[htbp]
\centering
\includegraphics[width=0.45\textwidth]{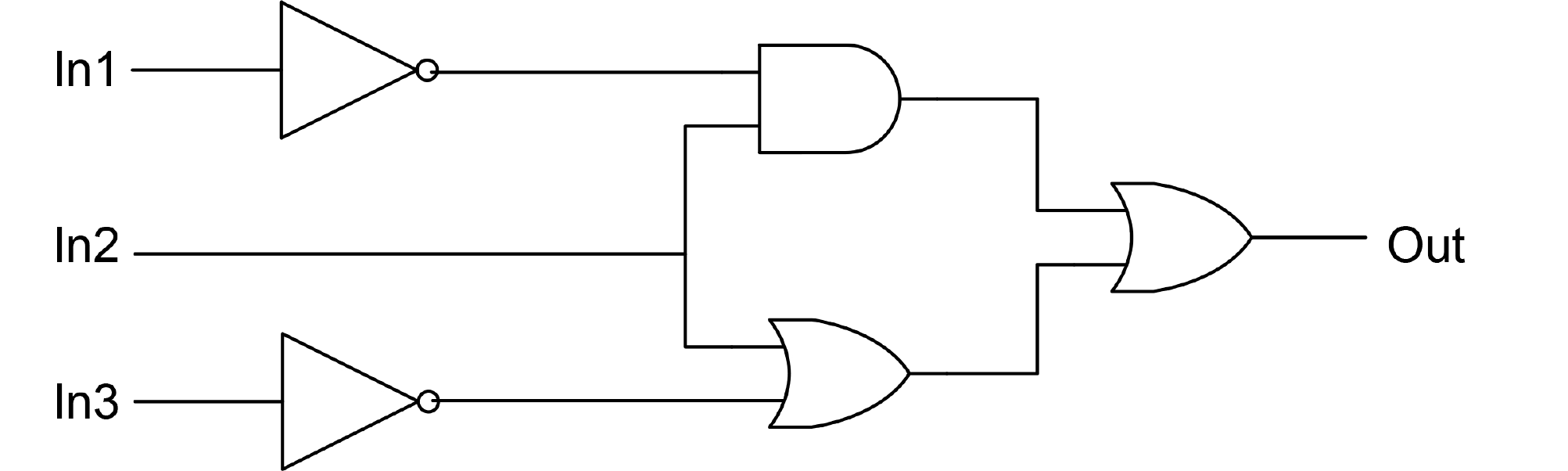}
\caption{Schematic design of second Boolean function \texttt{Out}= (\texttt{In1}' $\times$ \texttt{In2}) + ((\texttt{In3})' + (\texttt{In2})) with three (3) inputs, one (1) output and five (5) logic gates.}
\label{Fig5}
\end{figure}

\begin{figure}[htbp]
\centering
\includegraphics[width=0.45\textwidth]{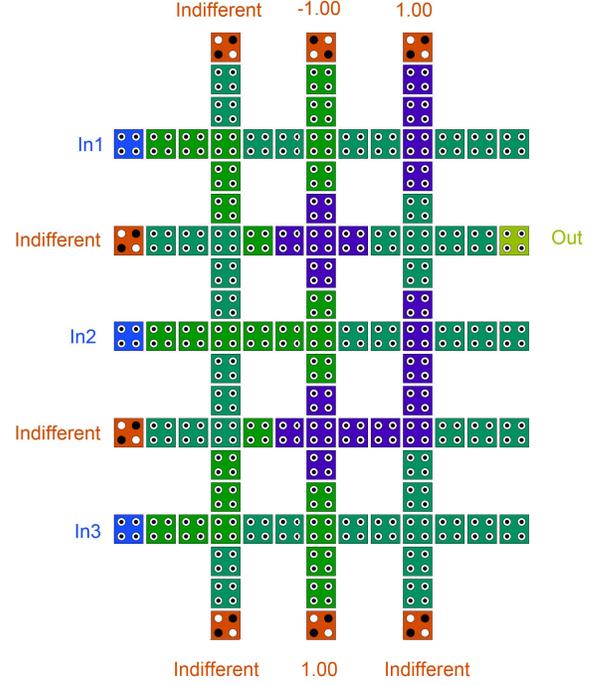}
\caption{The resulting crossbar QCA circuit corresponding to the Boolean function of Fig.~\ref{Fig5}.}
\label{Fig6}
\end{figure}

Fig.~\ref{Fig5} illustrates another simple Boolean function. The proposed function has no branches and consists of $3$ input and $1$ output signals. Following one more time the methodology described in Section~\ref{sec:method} it holds that $N_{inputs} \leq N_{lines} \leq N_{inputs} + N_{NOTgate}$ and so $3 \leq N_{lines} \leq 5$ and $N_{stages} \leq N_{columns} \leq N_{gates}$ and so $2 \leq N_{columns} \leq 5$. Fig.~\ref{Fig6} illustrates the corresponding QCA circuit. In this case, the designed QCA circuit consists of $ N_{lines} = 5$, due to the fact that each NOT gate of the circuit demands two lines in the crossbar to be implemented. The first line is occupied by the initial signal while the second one is occupied by the inverted one. It can also be observed that $N_{columns} = 3$, since the operation of the AND and of the first from the left OR gate presupposes the output from the NOT gates signals and hence, they can not constitute one stage. As a result, the two NOT gates as well as the AND and the first OR gate are incorporated into two different stages combined two by two respectively.
The timing of the circuit once again takes place according to the steps described earlier. The input cells of the \textit{input side} as well as the two NOT gates, are triggered by \texttt{Clock0}, that is presented with green color. The output signals of these gates with one of the input signals enter the AND and the first from the left OR gate and thus the operation clock of these gates is \texttt{Clock1} (purple color). The output signals resulting from the gates of the second stage enter the last OR gate and therefore the output signal is triggered by \texttt{Clock2}, represented by blue-turquoise color. Once more the indifferent or no longer needed cells are triggered by an appropriate clock, so they do not influence the functionality of the rest of the circuit. The results obtained by the implementation of the QCA circuit demonstrated in Fig. \ref{Fig6} using the QCADesigner are presented in Fig. \ref{Graphs}(b). Comparing to the previous circuit implementation, this time the clock under which the output $Out$ functions is \texttt{Clock2} and hence, the total delay of the circuit output corresponds to two clock phases. Once more, the extracted results are correct, revealing the method's stability and strength.

\begin{figure}[htbp]
\centering
\includegraphics[width=0.45\textwidth]{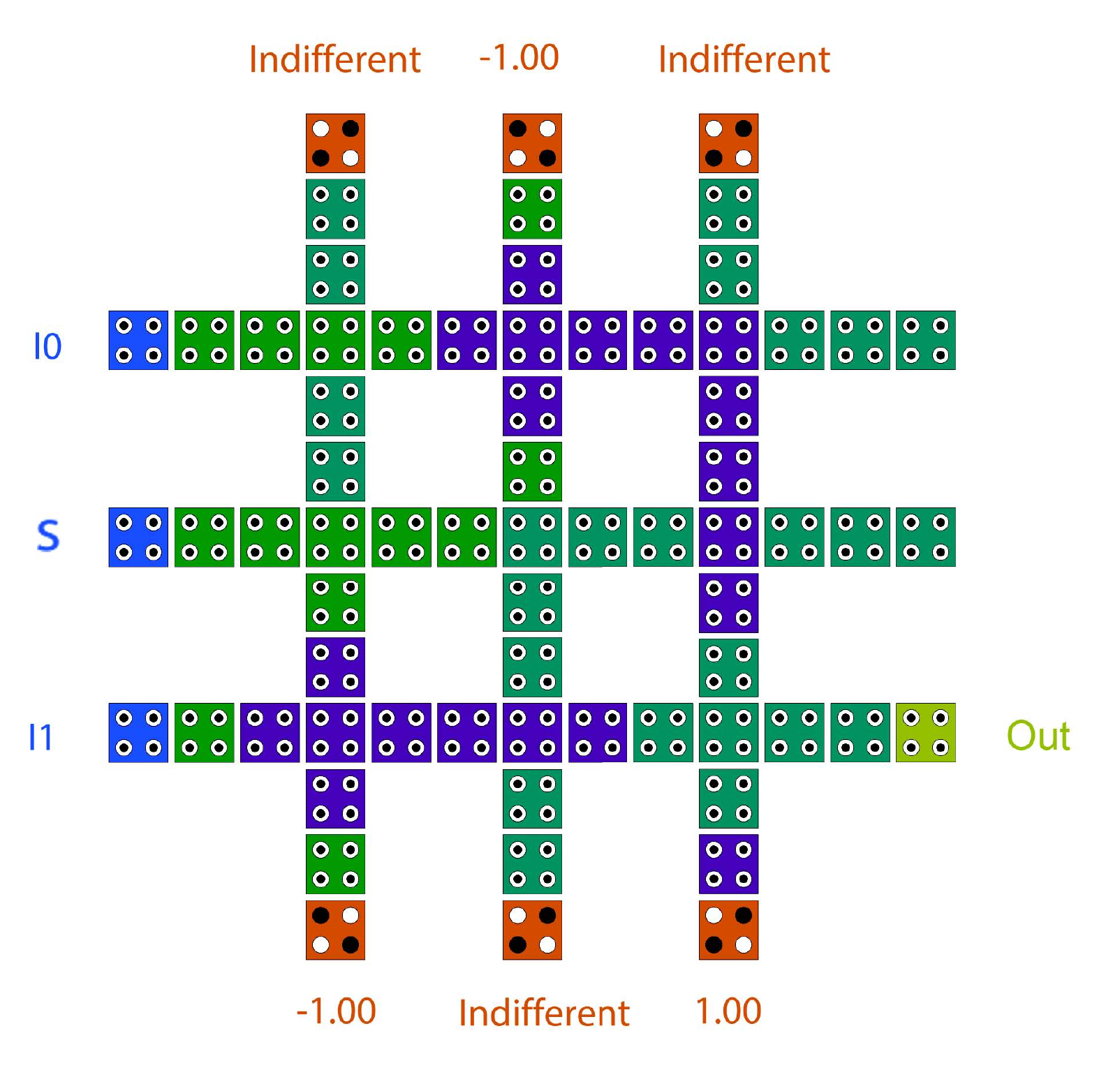}
\caption{The resulting crossbar QCA circuit corresponding to the 2-to-1 multiplexer.}
\label{Mux21QCA}
\end{figure}

\begin{figure}[htbp]
\centering
\includegraphics[width=0.45\textwidth]{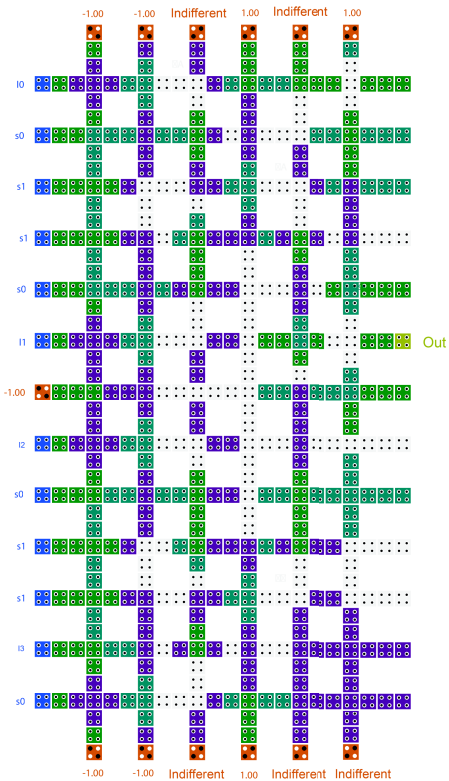}
\caption{The resulting crossbar QCA circuit of the 4-to-1 multiplexer.}
\label{Mux41QCA}
\end{figure}

Then, two other digital circuits were designed and simulated according to the proposed methodology: the 2-to-1 and the 4-to-1 multiplexer. The QCA implementation of the former is demonstrated in Fig. \ref{Mux21QCA}. The crossbar of the implemented circuit consists of $N_{lines} = 3$ and $N_{columns} = 3$, while the timing of all cells obeys all the rules of the proposed method. As far as the design of the 4-to-1 multiplexer is concerned, a more generalized form of the proposed method is needed. As already mentioned in the previous Section, in order for a circuit that contains more than $2$ AND or OR gates at the same stage to be designed in the crossbar architecture, it is demanding to slightly modify the design rule regarding the program cells: in this case the constant values needed for the operation of the AND or OR gate are able to be inserted to the circuit from the \textit{input side} and not only from the \textit{program lines}. Based on this relaxation of the design rules, any complex digital circuit (without any limitations) is able to be successfully designed in a crossbar by simple following the proposed design steps. Finally, the logical QCA implementation of the 4-to-1 multiplexer is demonstrated in Fig. \ref{Mux41QCA}. The simulation results of the two pre-mentioned multiplexers are depicted in Fig. \ref{Graphs}(c) and Fig. \ref{Graphs}(d), respectively. 

\begin{figure}[htbp]
\centering
\includegraphics[width=0.45\textwidth]{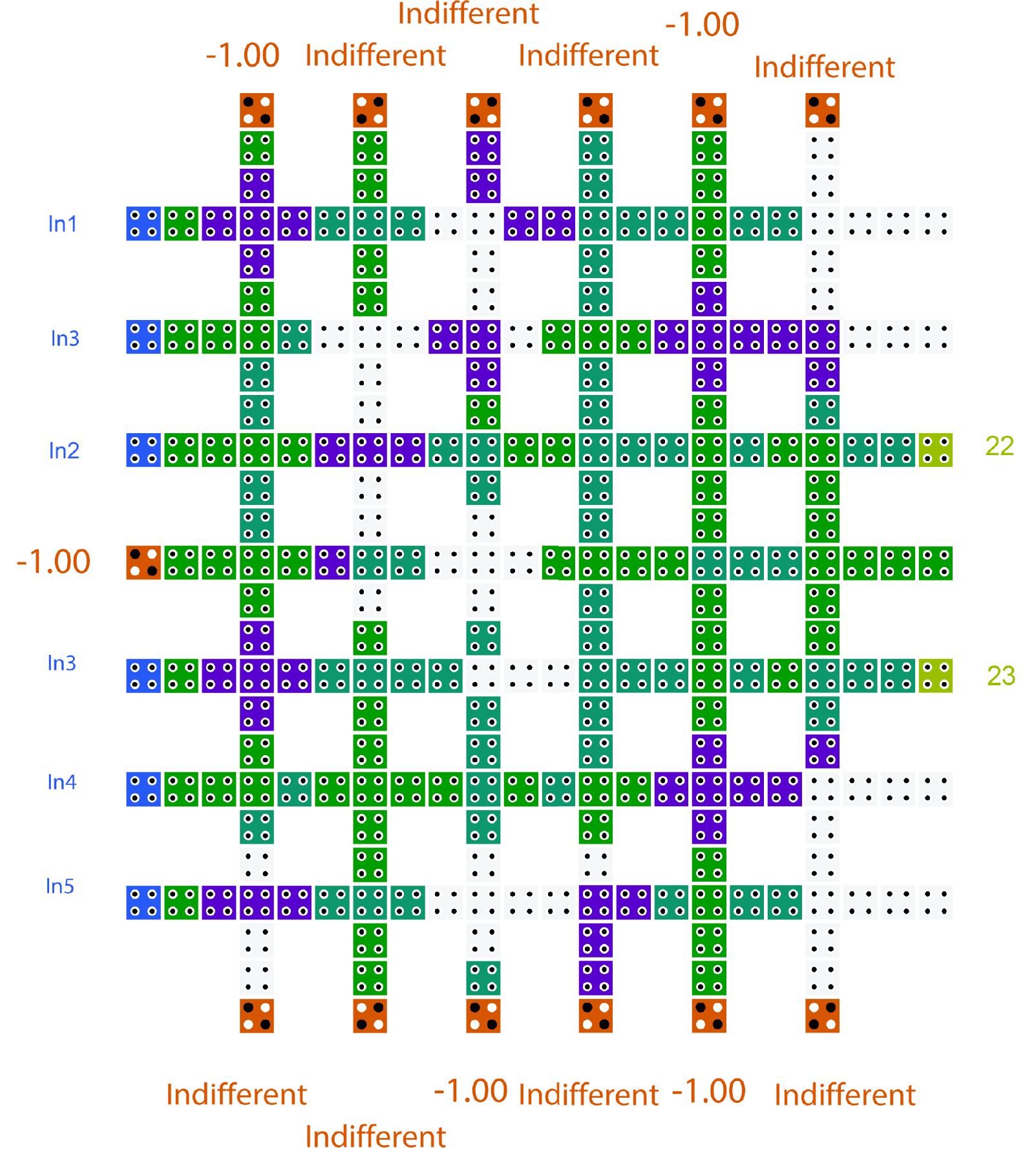}
\caption{The resulting crossbar QCA circuit of the ISCAS'85 c17 circuit.}
\label{c17QCA}
\end{figure}

\begin{figure}[htbp]
\centering
\includegraphics[width=0.45\textwidth]{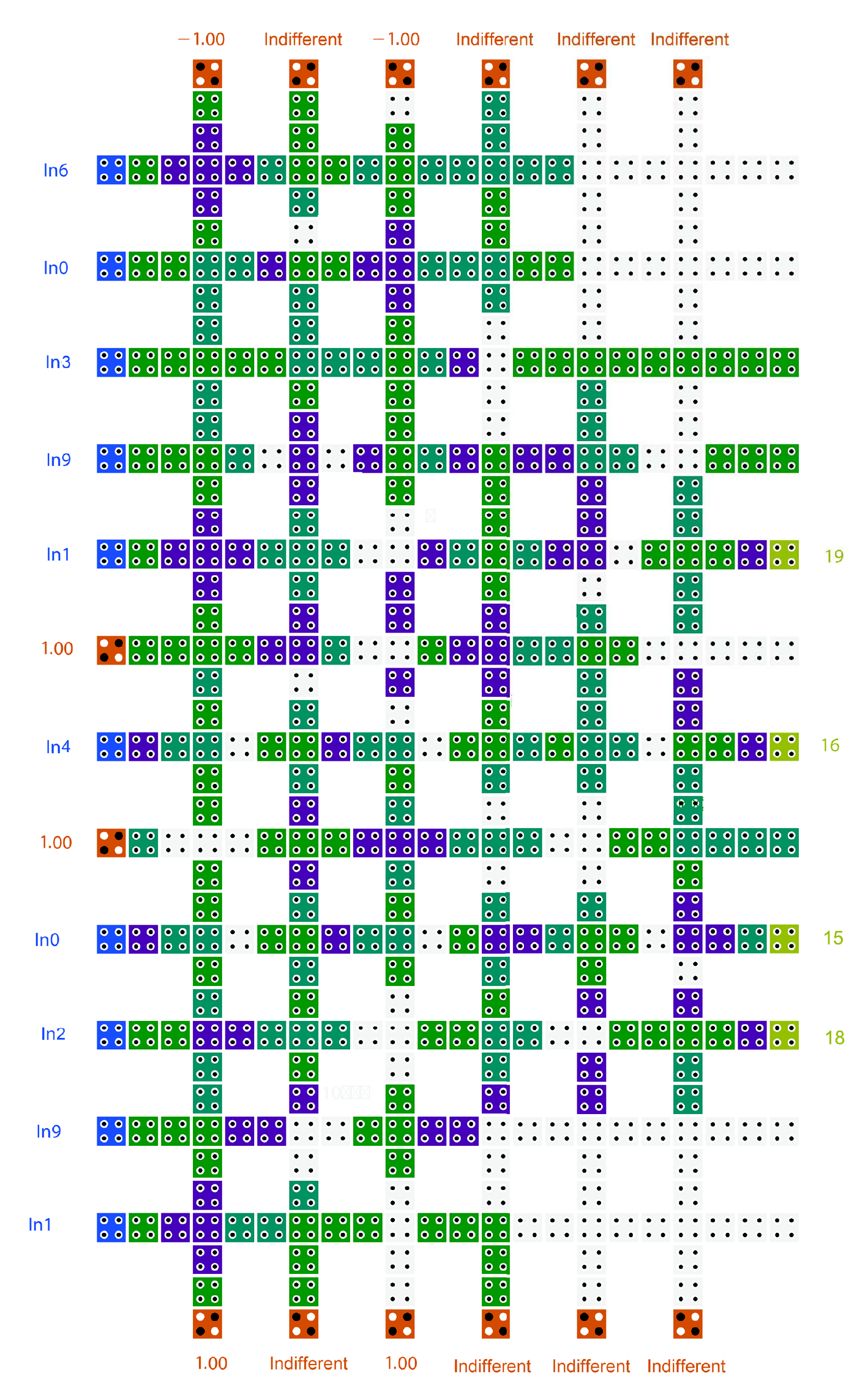}
\caption{The resulting crossbar QCA circuit of the ISCAS'89 s27 circuit.}
\label{s27QCA}
\end{figure}

The next design experiments comprise the design and simulation of two ISCAS benchmark circuits \cite{785838}: the combinational ISCAS'85 c17 as well as the sequential ISCAS'89 s27 circuits \cite{flanigan2009scalable}.
As it is easily noticed, both the circuits contain branches and therefore, their successful design presupposes, according to the proposed method, the transformation of each logic circuit to one without any branches; an efficient tactic to achieve the aforementioned is, simply, the multi-insertion of one or more input signals to the circuit. As in the case of the 4-to-1 multiplexer QCA design, both these circuits demand the implementation of more than one AND or OR gate in a single stage. Hence, following the same logic as before, in both QCA implementations of the ISCAS circuits (as shown in  Fig. \ref{c17QCA} and  Fig. \ref{s27QCA}) an appropriate number of constant values is inserted to the circuit from the \textit{input side}. The QCAdesigner simulation results, demonstrated in Fig. \ref{Graphs}(e) and Fig. \ref{Graphs}(f) respectively, indicate the applicability of the proposed method to any large and complex circuits and showed stable operation in both cases.

\begin{figure}[htbp]
\centering
\includegraphics[width=0.45\textwidth]{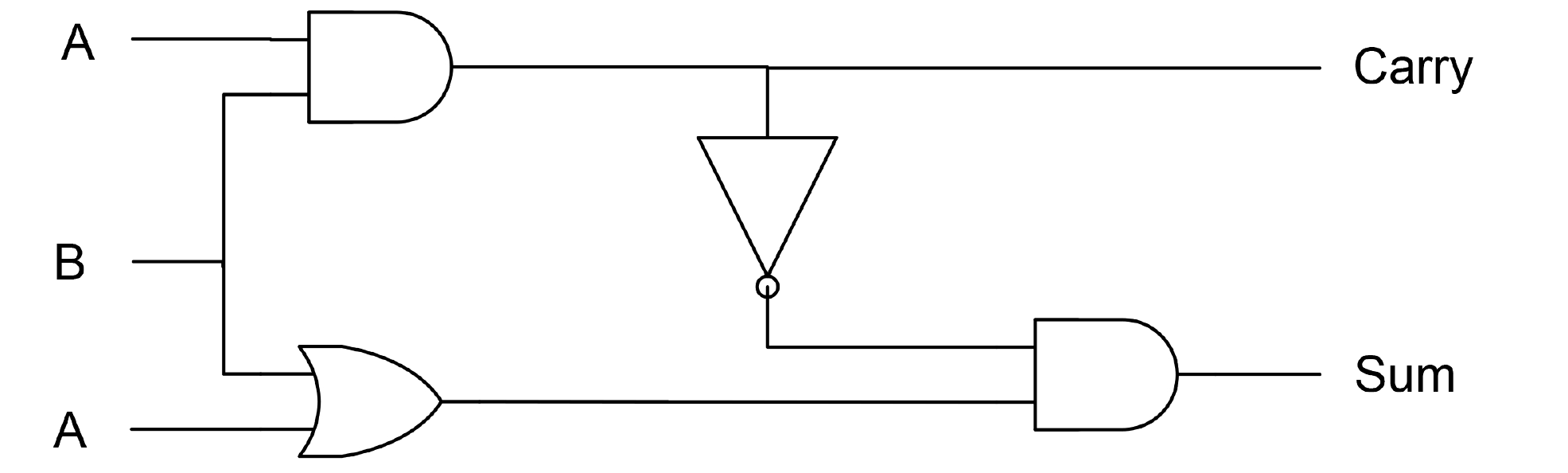}
\caption{The logic circuit of the Half Adder.}
\label{Fig7}
\end{figure}

\begin{figure}[htbp]
\centering
\includegraphics[width=0.45\textwidth]{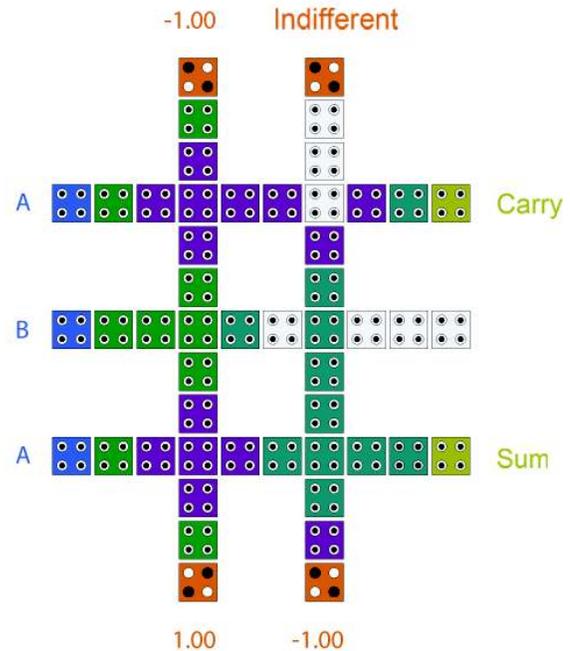}
\caption{The crossbar QCA circuit of the 1-bit Half Adder.}
\label{Fig8}
\end{figure}

The methodology is now applied to the design of the half-adder. As it is known, the half-adder circuit consists of an AND and a XOR gate. Taking advantage of the proposed programmable Boolean set, namely AND, OR and NOT, described earlier, all other gates are decomposed into the aforementioned gates without any further loss of generality. In the method introduced in this paper, each circuit under construction should contain only the basic gates, namely the AND, OR and NOT gates. Therefore, as described in previous section all other gates are decomposed into the basic gates in order for the method to be applicable. Thus, we use an implementation of the half-adder based on the provided Boolean set, as shown in Fig. \ref{Fig7}. Fig. \ref{Fig8} illustrates the QCA implementation of the half-adder circuit. Following the same procedure as before the QCA circuit consists of $ N_{lines} = 3$ and $N_{columns} = 2$, both corresponding, according to the proposed method, to the minimum available numbers of lines and columns. As far as the timing of the useful as well as of the indifferent cells of the QCA circuit is concerned, the procedure for the construction of the desired QCA circuit does not present any differences from the one presented already in the two previous designs. The output results, i.e. $Carry$ and $Sum$, for all the possible value pairs of the inputs $A$ and $B$, as they were generated by the proposed QCA implementation of the half-adder circuit, are presented in Fig. \ref{Graphs}(g).

\begin{figure}[htbp]
\centering
\includegraphics[width=0.45\textwidth]{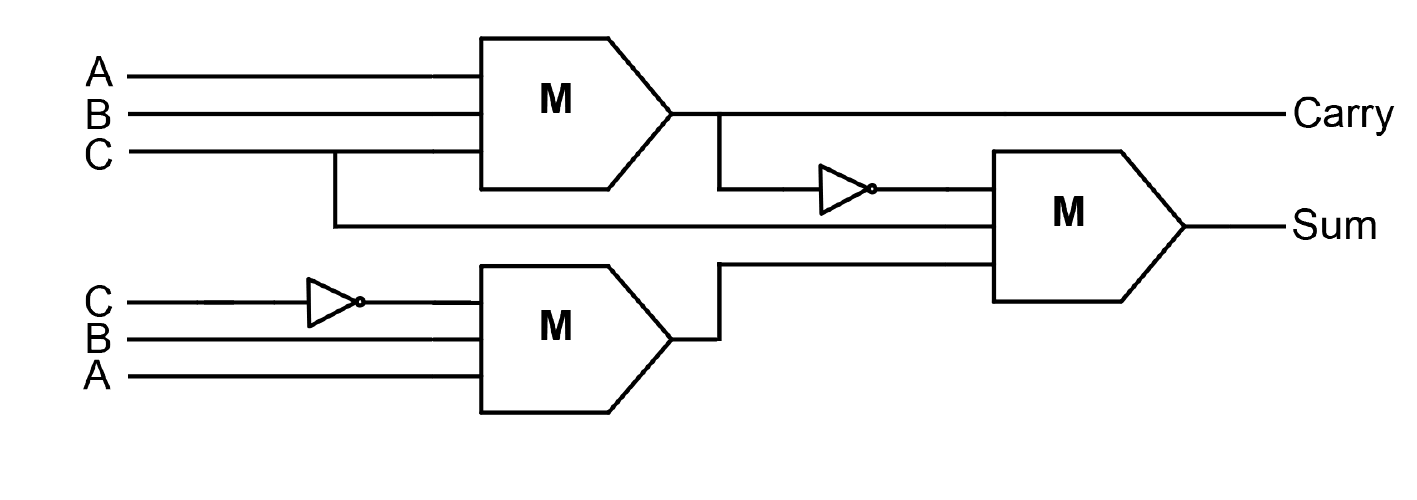}
\caption{The logic circuit of the 1-bit Full Adder, using the Majority logic gate.}
\label{Fig9}
\end{figure}

\begin{figure}[htbp]
\centering
\includegraphics[width=0.45\textwidth]{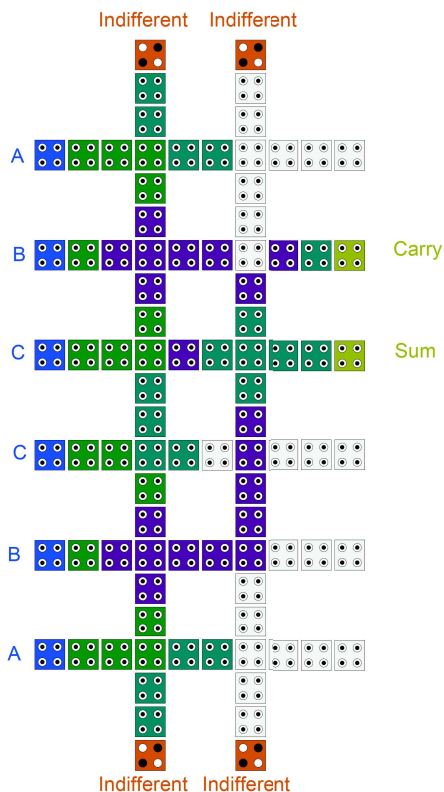}
\caption{The crossbar QCA circuit of the 1-bit Full Adder.}
\label{Fig10}
\end{figure}

\begin{figure*}[htbp]
\begin{center}
\begin{tabular}{c@{}c@{}c@{}}
\includegraphics[width=0.4\textwidth ,height=150pt]{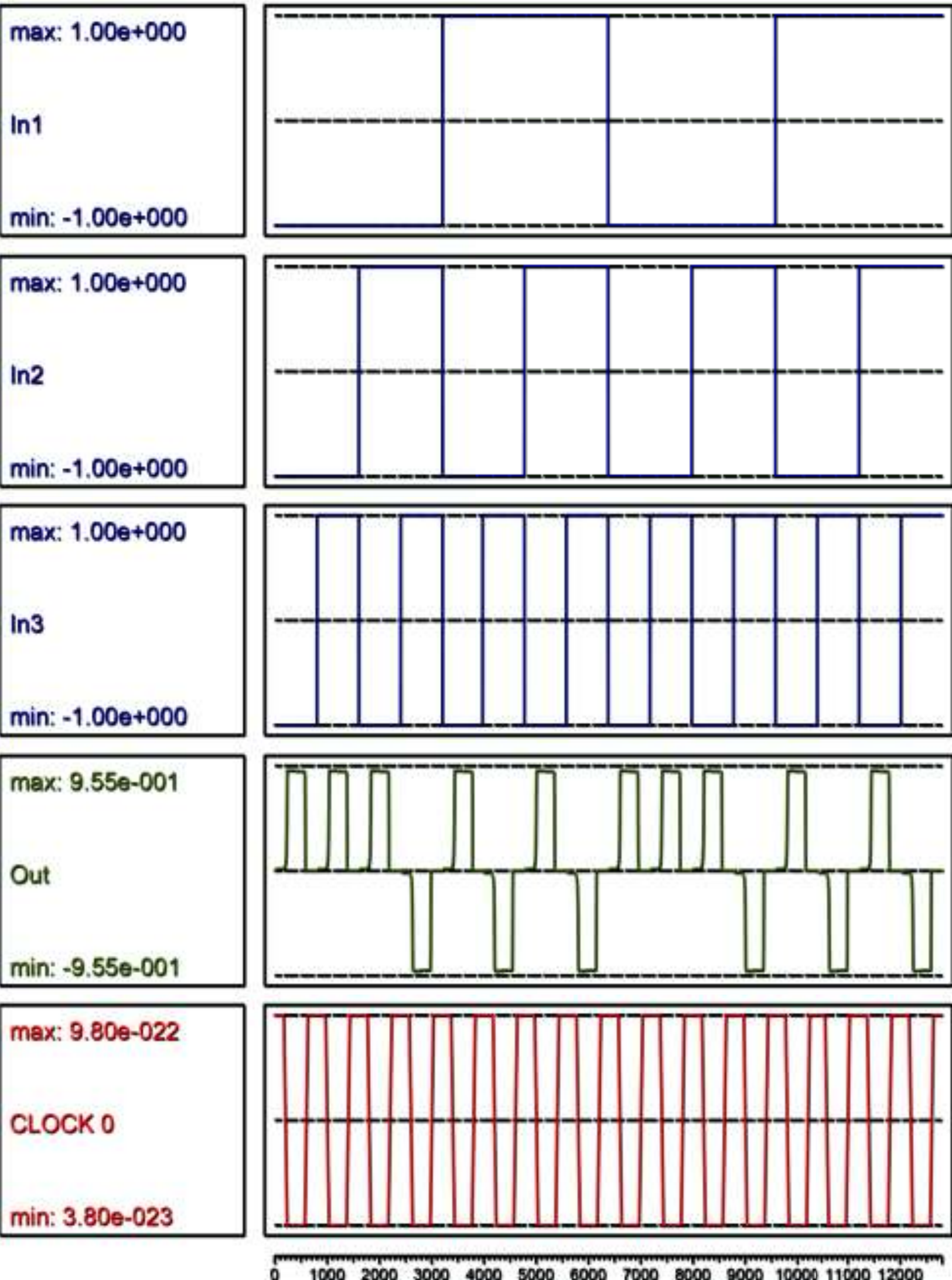}&
\textit{ }\textit{ }\textit{ }\textit{ }\textit{ } &
\includegraphics[width=0.4\textwidth ,height=150pt]{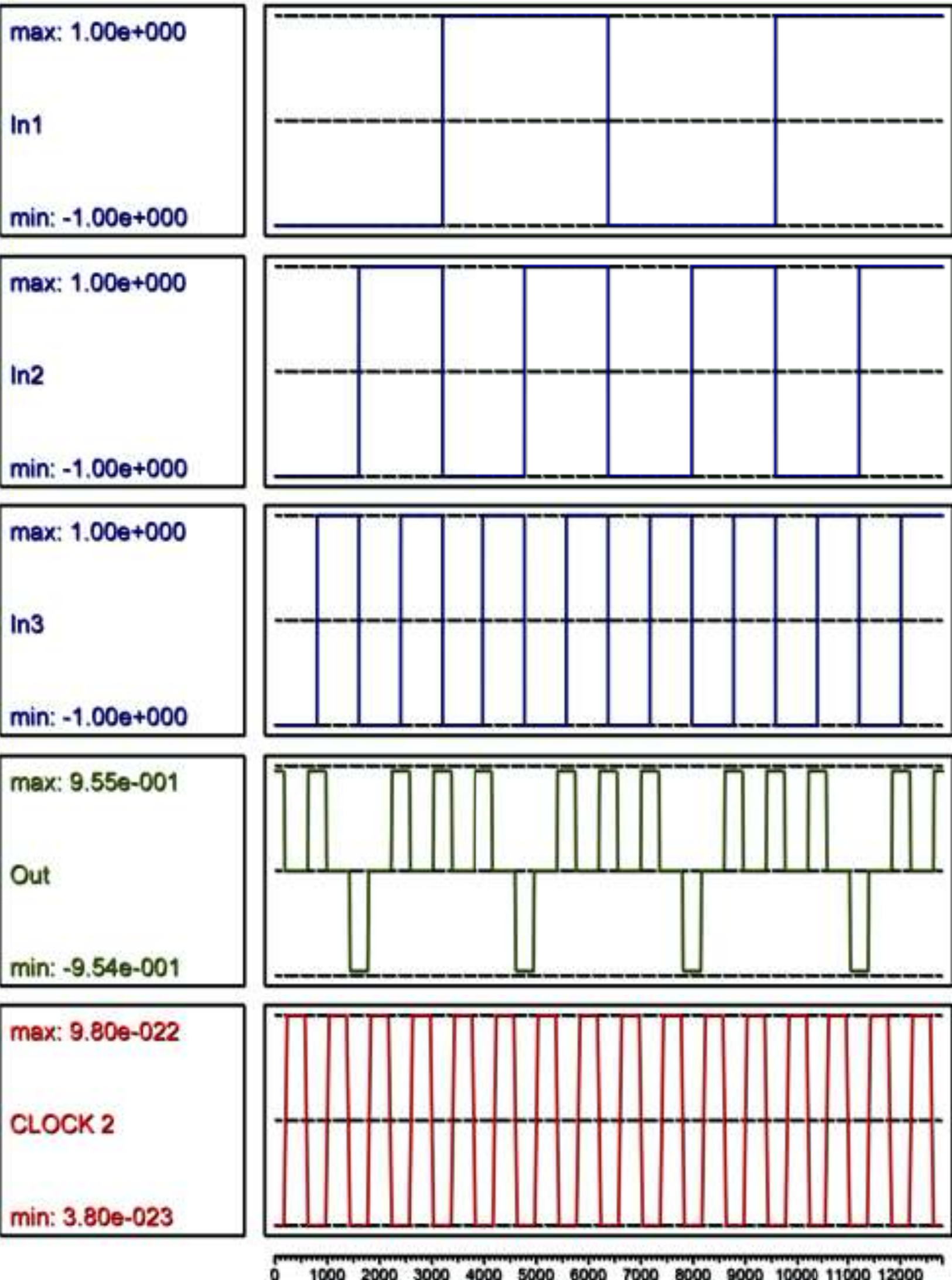}\\
{\small(a)} &  & {\small(b)}\\
\includegraphics[width=0.4\textwidth ,height=150pt]{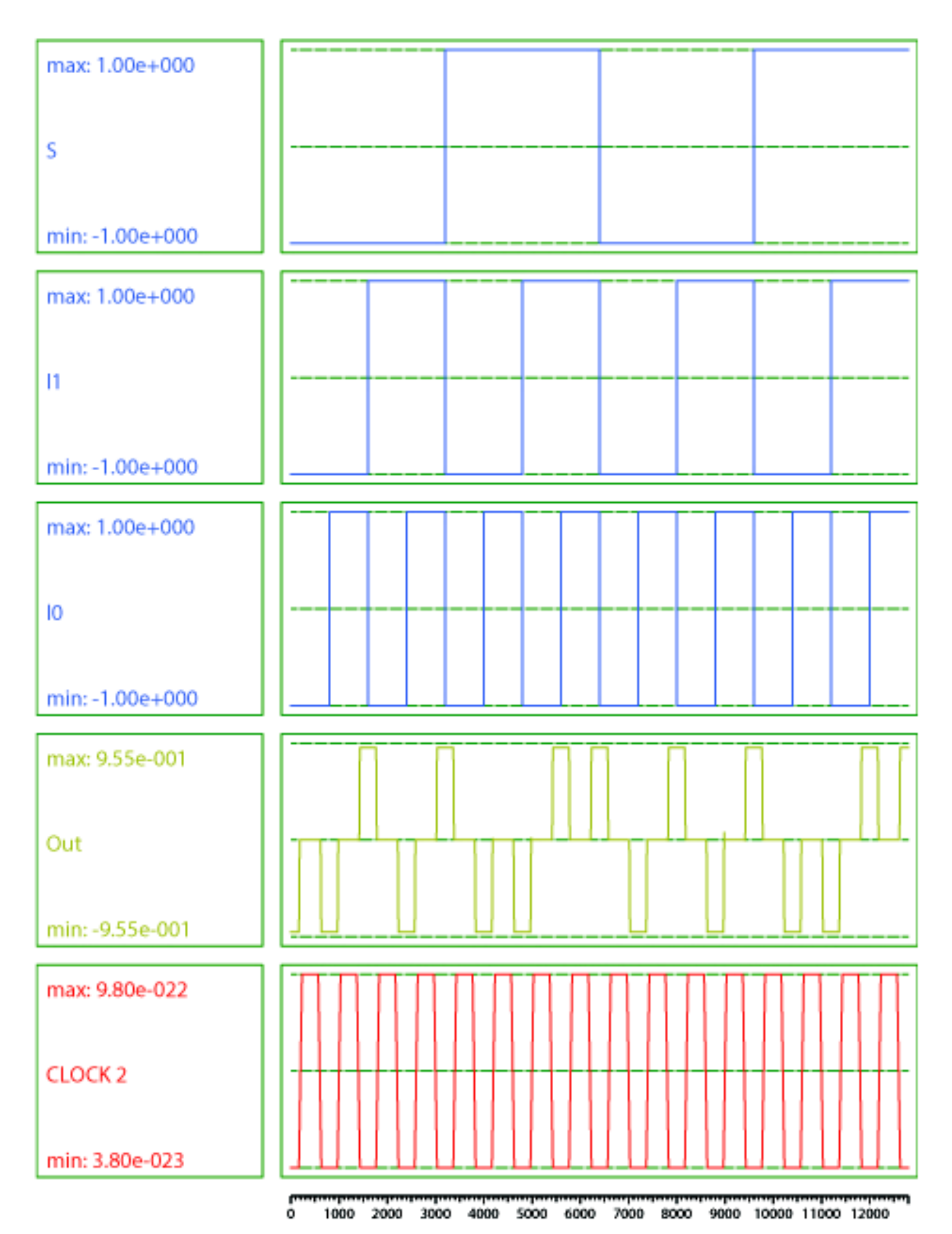}&
\textit{ }\textit{ }\textit{ }\textit{ }\textit{ } &
\includegraphics[width=0.4\textwidth ,height=150pt]{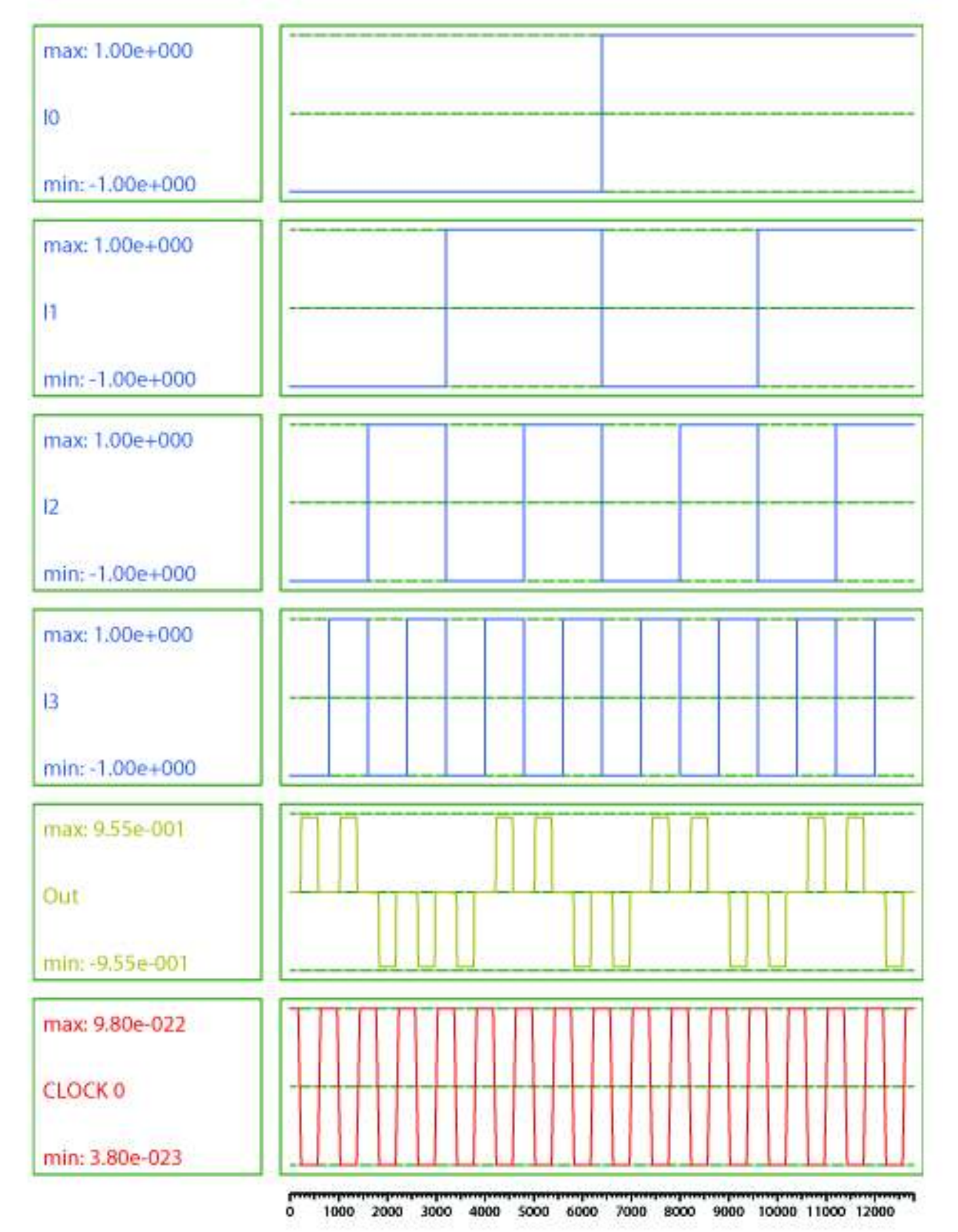}\\
{\small(c)} &  & {\small(d)}\\
\includegraphics[width=0.4\textwidth ,height=150pt]{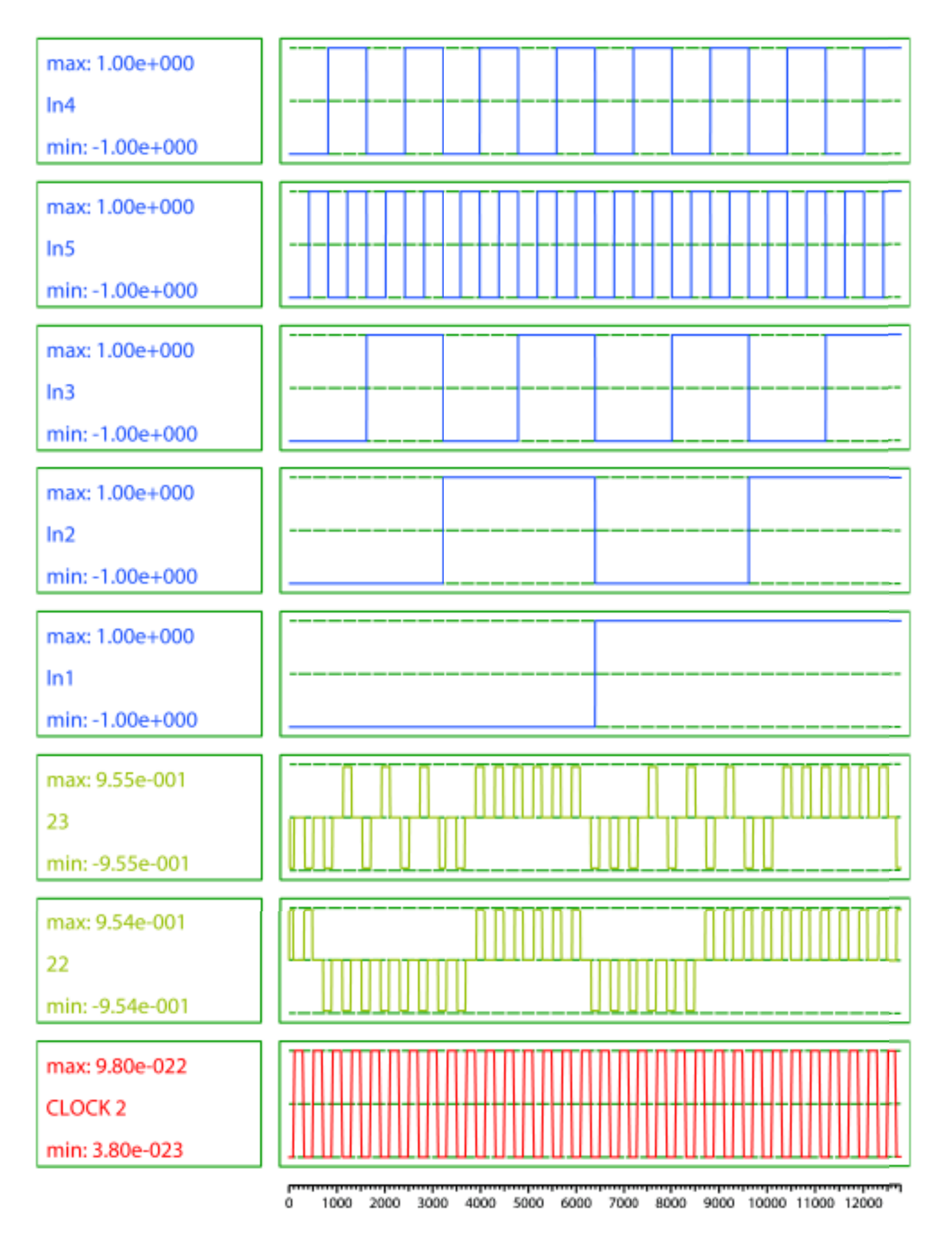}&
\textit{ }\textit{ }\textit{ }\textit{ }\textit{ } &
\includegraphics[width=0.4\textwidth ,height=150pt]{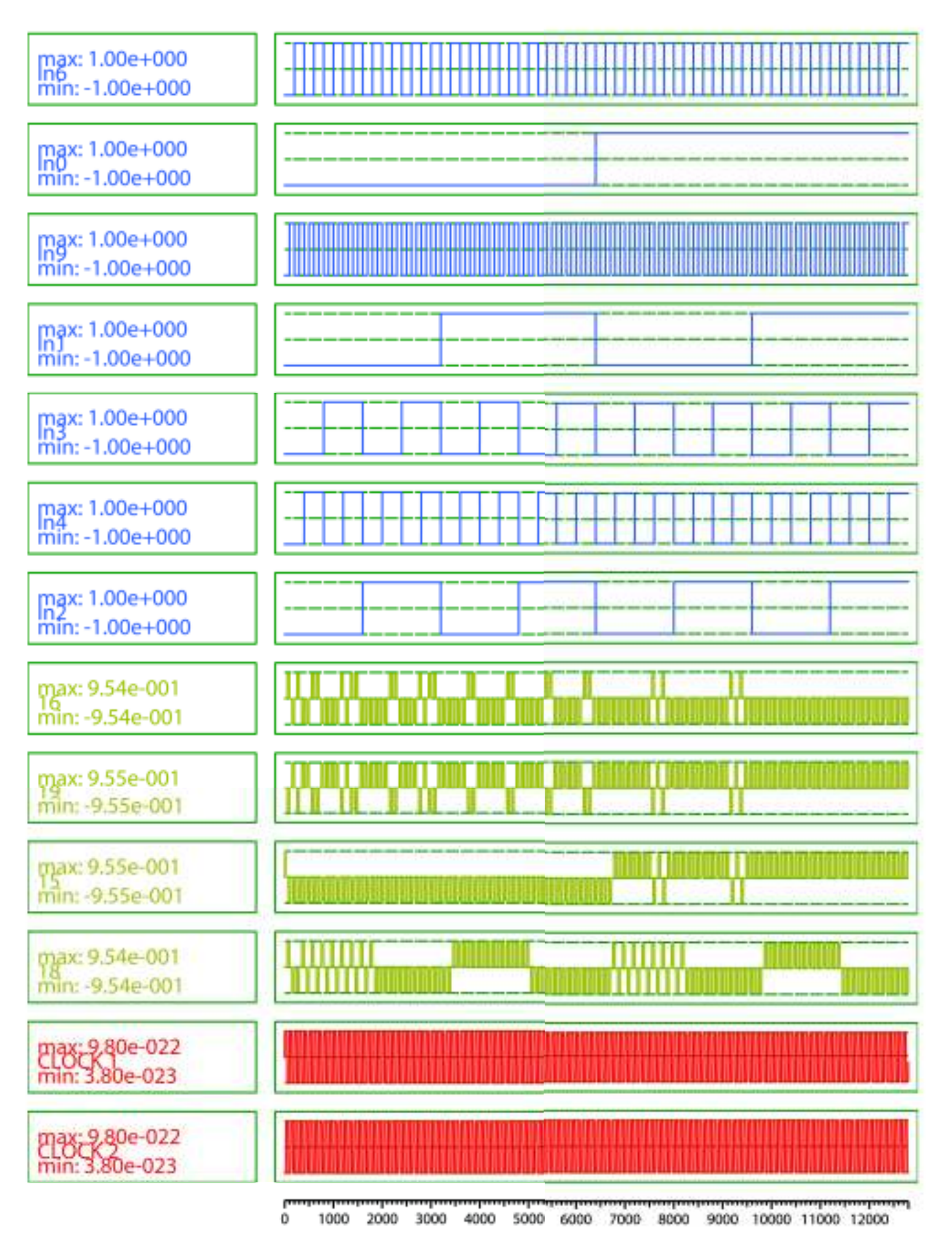}\\
{\small(e)} &  & {\small(f)}\\
\includegraphics[width=0.4\textwidth ,height=150pt]{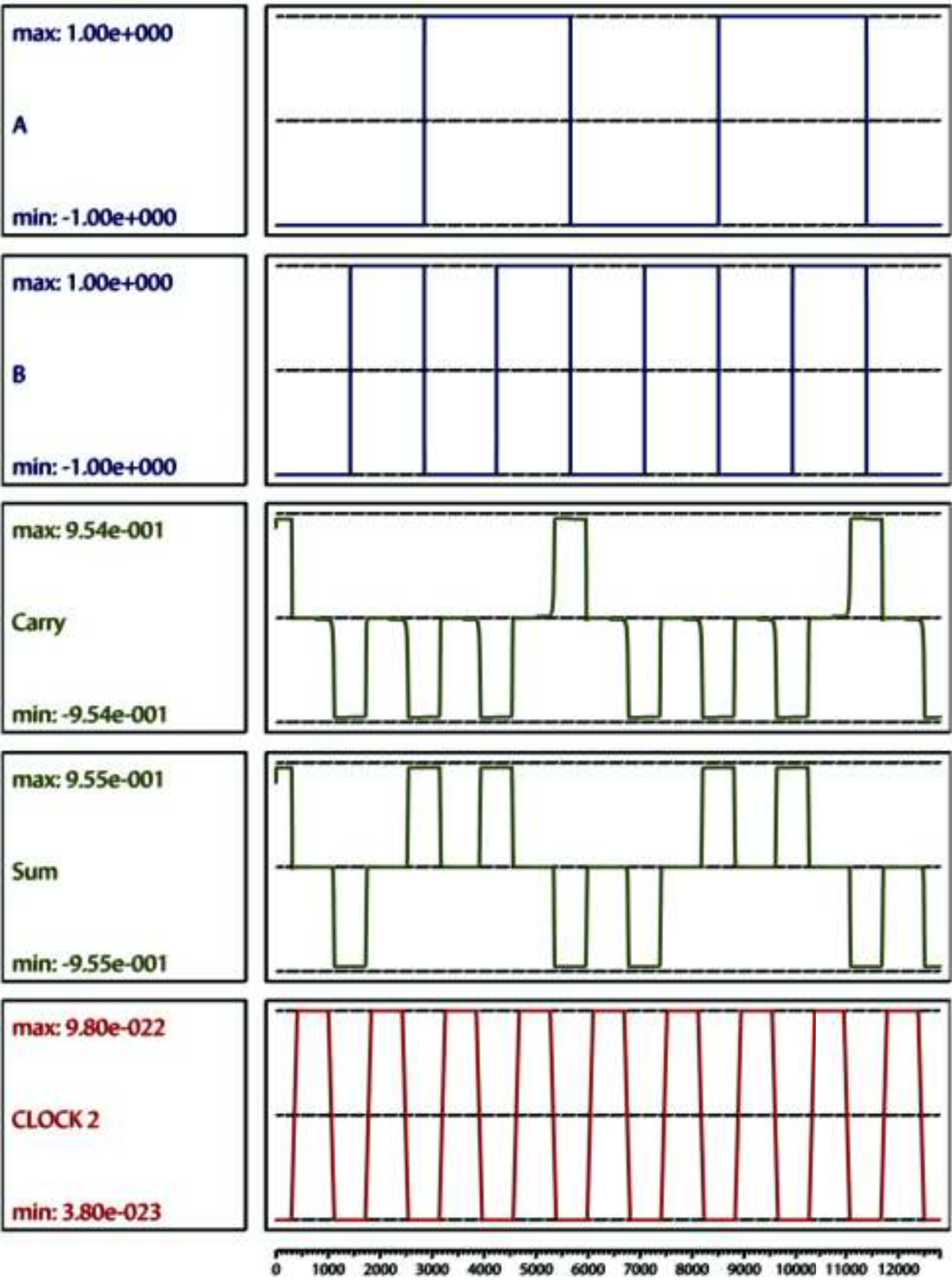}&
\textit{ }\textit{ }\textit{ }\textit{ }\textit{ } &
\includegraphics[width=0.4\textwidth ,height=150pt]{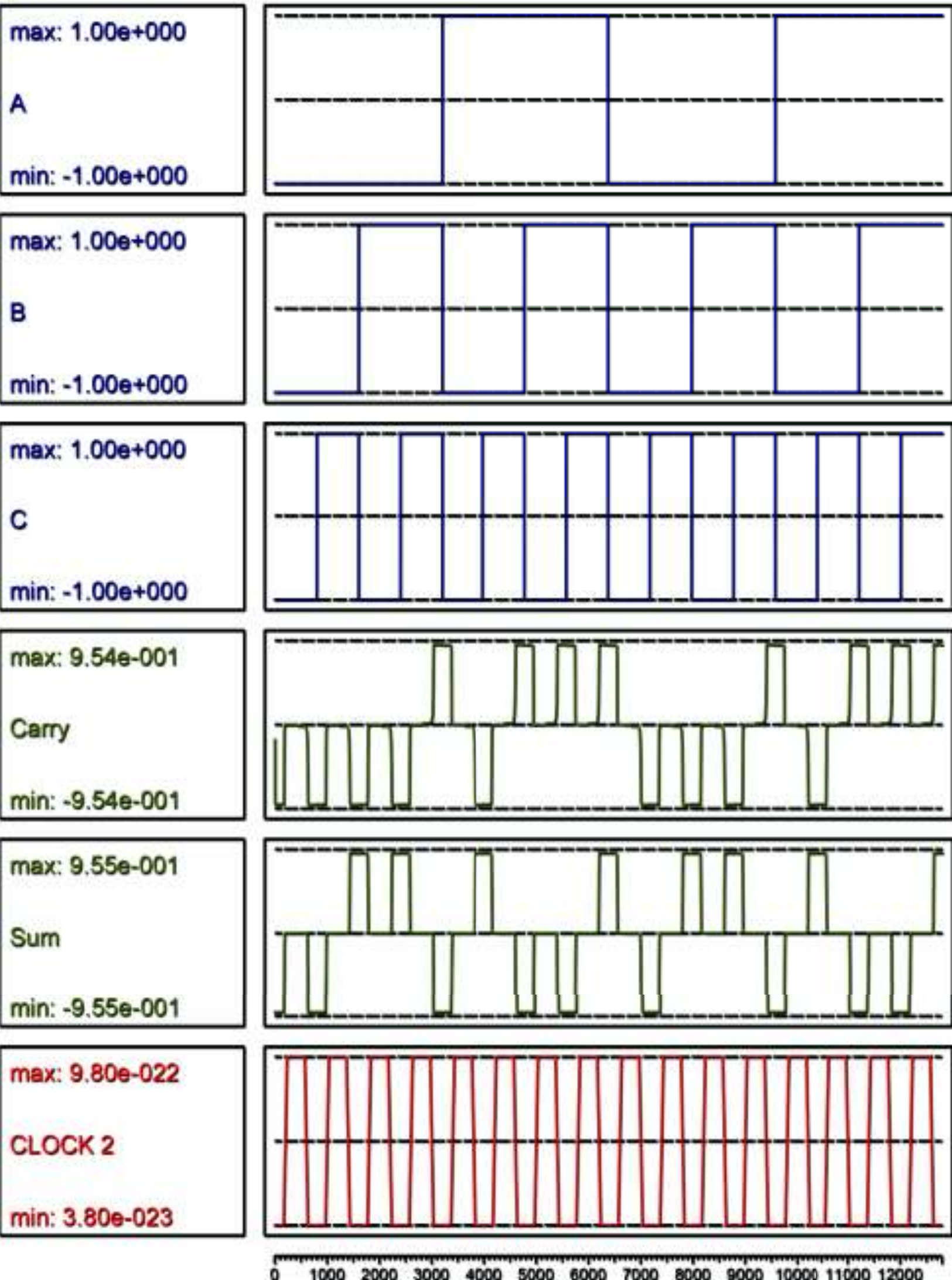}\\
{\small(g)} &  & {\small(h)}
\end{tabular}
\end{center}
\caption{QCADesigner simulation results for all circuits treated: (a) the first QCA Circuit demonstrated in Fig. \ref{Fig4}, (b) the second QCA Circuit demonstrated in Fig. \ref{Fig6}, (c) the 2-to-1 multiplexer of Fig. \ref{Mux21QCA}, (d) the 4-to-1 multiplexer of Fig. \ref{Mux41QCA}, (e) the benchmark c17 of the ISCAS'85 collection illustrated in Fig. \ref{c17QCA}, (f) the benchmark s27 of the ISCAS'89 collection depicted in Fig. \ref{s27QCA}, (g) the 1-bit QCA half adder of Fig. \ref{Fig8} and (h) the 1-bit QCA full adder of Fig. \ref{Fig10}. For each circuit, the input (blue graphs) and the output (yellow graphs) signals as well as the clock (red graphs) under which the outputs operate are shown.}
\label{Graphs}
\end{figure*}

The logic circuit of a full-adder using Majority gates is shown in Fig.~\ref{Fig9}. We choose to use this logic circuit in order to display the flexibility of the method. In more details, the circuit contains $3$ Majority gates and $2$ inverters, whereas all $3$ input signals of the 1-bit full-adder are inserted into the circuit twice as presented. 
The resulting QCA circuit using the crossbar architecture is depicted in Fig.~\ref{Fig10}. The constructed crossbar has $6$ lines and only $2$ columns. At this point, it should be mentioned that the \textit{input side} of the circuit contains all $3$ input signals twice, since there is no need using more lines for the operation of the NOT gates. As far as the columns of the QCA circuit are concerned, these are equal to the number of stages of the circuit, since in the first column/ stage the operation of the $2$ Majority and $1$ NOT gate takes place, whereas the second one consists of the third Majority and the other NOT gate. The clocking strategy followed to trigger each cell is the one described both in this and in the previous section. The results shown in Fig. \ref{Graphs}(h) evidence the proper and stable operation of the proposed 1-bit QCA full-adder design.

\begin{table*}[!t]
\renewcommand{\arraystretch}{1.3}
\caption{Analysis of the implemented QCA circuits in terms of area consumption, cell count and clock phases.}
\label{table2}
	\centering
		\begin{tabular}
{ c  c  c  c }
			\toprule
			\bfseries \small{Circuit} & \bfseries \small{Approximated Area ($\mu m^2$)} & \bfseries \small{Cell Count} & \bfseries \small{Delay (clock phases)}\\
			 			\toprule
			 			
\small{Circuit of Fig. \ref{Fig3}} & \small{0.0676} & \small{69} & \small{4} \\
												
           \small{Circuit of Fig. \ref{Fig5}} & \small{0.0988} & \small{107} & \small{3} \\

           \small{Mux 2-to-1} & \small{0.0676} & \small{69} & \small{3} \\

           \small{Mux 4-to-2} & \small{0.3784} & \small{466} & \small{13} \\
	
           \small{ISCAS'85 c17} & \small{0.22} & \small{262} & \small{7} \\
           
           \small{ISCAS'89 s27} & \small{0.352} & \small{432} & \small{15}  \\

           \small{1-bit Half Adder} & \small{0.052} & \small{50} & \small{3} \\

			 			\small{1-bit Full Adder} & \small{0.087} & \small{92} & \small{3} \\

           \bottomrule
           
		\end{tabular}	
\end{table*}

\begin{table*}[!t]
\renewcommand{\arraystretch}{1.3}
\caption{Comparison of 1-bit Full Adder QCA designs.}
\label{table1}
	\centering
		\begin{tabular}
{ c  c  c  c  c }
			\toprule
			\bfseries \small{References} & \bfseries \small{Approximated Area ($\mu m^2$)} & \bfseries \small{Cell Count} & \bfseries \small{Delay (clock phases)} & \bfseries \small{Wire Crossings}\\
			 			\toprule
			 			
\small{\cite{Cho07}} & \small{0.174} & \small{135} & \small{5} &  \\
												
           \small{\cite{Cho09}} & \small{0.092} & \small{86} & \small{3} & \\

           \small{\cite{hashemi2012efficient}} & \small{0.051} & \small{79} & \small{5} & Multi-layer\\

           \small{\cite{navi2010new}} & \small{0.039} & \small{73} & \small{3} & \\
	
           \small{\cite{hayati2012design}} & \small{0.031} & \small{44} & \small{4} & \\
           \midrule
           \small{\cite{kim2007robust}} & \small{0.362} & \small{220} & \small{12}  & \\

           \small{\cite{Tougaw94}} & \small{0.198} & \small{192} & \small{Not applicable} & \\

			 			\small{\cite{zhang2004method}} & \small{0.157} & \small{145} & \small{5} & Coplanar\\

           \small{\cite{hanninen2010binary}} & \small{0.098} & \small{101} & \small{8} & \\

      \small{Proposed} & \bfseries \small{0.087} & \bfseries \small{92} & \bfseries \small{3} & \\
           \bottomrule
           
		\end{tabular}	
\end{table*}

Using QCADesigner, complexity, delay and area consumption of QCA circuits can easily be obtained \cite{citeulike:344490}. Table \ref{table2} demonstrates the corresponding implementation details (approximated area, number of cells and clock phases) of all the eight QCA circuits designed using the proposed methodology. For example, in case of 1-bit full adder, the resulting QCA crossbar is implemented using $92$ QCA cells, takes only $3$ clock phases ($0.75$ clock cycles) to produce the desired output signals and occupies approximately $0.087 um^2$. 

What is more, the proposed QCA implementation of the 1-bit full adder is compared with other QCA 1-bit full adders found in literature \cite{Tougaw94,Cho07,zhang2004method,kim2007robust,Cho09,navi2010new,hanninen2010binary,hayati2012design,hashemi2012efficient}. The effectiveness and superiority of the proposed QCA 1-bit full adder implementation in terms of power delay, area and number of cells can be easily acknowledged as shown in Table \ref{table1}. Please notice that in the designs found in \cite{Cho07,Cho09,hashemi2012efficient,navi2010new,hayati2012design} the provided circuits result from multi-layer crossover designs. Unfortunately, till now there are serious questions about how these multi-layer crossover designs can be realized in practice, since they require two overlapping active layers with via connections \cite{Cho:2007:SPM:1270377.1270444}.

\section{Conclusions}
\label{sec:conclusions}

Although QCA is a nanoelectronic technology recognized as a one of the top emerging technologies, its issues have not been adequately addressed and therefore, there is no consistent or standard framework in designing QCA circuits. Thus, in this paper, a novel design methodology aiming at implementing Boolean logic functions using programmable crossbar QCA circuits is presented. As an additional contribution, in the context of this work, a novel implementation of the QCA NOT gate is introduced facilitating the combination of the QCA nanotechnology with the crossbar architecture. This means that, the programmable AND and OR QCA gates in conjunction with the proposed QCA inverter are able to be implemented in a cross point of the crossbar and hence, to form a universal Boolean set suitable for designing any Boolean logic QCA circuit and implementing general computation.

In particular, we provide the QCA circuits' designer with a unified methodology based on specific rule steps associated with (a) the proper dimension of  the crossbar QCA circuit, (b) the appropriate timing of cells and (c) the exact programming and activation of the gates. The timing together with the programming issues of QCA cells and gates are successfully handled in every case with a cascadable easy to follow way. Following the proposed strategy, the QCA design of any Boolean logic circuit in a crossbar can be successfully achieved.

The proposed approach was applied to several Boolean logic circuits in order to evaluate its effectiveness; in this paper, $8$ of them were presented: two simple ones, the 2-to-1 and 4-to-1 multiplexers, the ISCAS'85 c17 and ISCAS'89 s27 benchmark circuits, the 1-bit QCA half adder and, finally, the 1-bit QCA full adder. The corresponding QCA crossbar circuits were constructed according to the proposed methodology and were designed and simulated using the standard QCA computer-aided (CAD) tool QCADesigner \cite{citeulike:344490}. The obtained results reveal the effectiveness, the stability and the reliability of our method not to mention the fact that the designed QCA circuits provide significant hardware savings comparing to other examined conventional QCA designs. This is evident, due to the fact that our 1-bit QCA full adder outperforms in terms of circuit area consumption, latency as well as complexity all the QCA full adders with coplanar wire crossings published in the literature.

\ifCLASSOPTIONcaptionsoff
  \newpage
\fi

\bibliographystyle{IEEEtran}
\bibliography{QCA}

\end{document}